\documentclass[manuscript]{aastex}

\usepackage{amsmath}
\usepackage{amssymb}
\usepackage{graphicx}
\usepackage{epstopdf}

\shorttitle{L1251-C}
\shortauthors{Kim et al.}

\begin{document}

\newcommand{\hcop}{HCO$^+$ 1--0}
\newcommand{\htcop}{H$^{13}$CO$^+$ 1--0}
\newcommand{\nhp}{N$_2$H$^+$ 1--0}
\newcommand{\water}{H$_2$O}
\newcommand{\htco}{H$_2$CO 2$_{12}$--1$_{11}$}
\newcommand{\tco}{CO 1--0}
\newcommand{\hcn}{HCN 1--0}
\newcommand{\m}{\micron}
\newcommand{\tcot}{CO 2--1}
\newcommand{\thcot}{$^{13}$CO 2--1}
\newcommand{\kms}{km s$^{-1}$}
\newcommand{\vlsr}{$v_{\textrm{lsr}}$}

\title{Infrared and Radio observations of a small group of protostellar objects in the molecular core, L1251-C}

\author{Jungha Kim$^1$, Jeong-Eun Lee$^{1,4}$, Minho Choi$^2$, Tyler L. Bourke$^3$, Neal J. Evans II$^4$, James Di Francesco$^5$, Lucas A. Cieza$^{6}$, Michael M. Dunham$^7$, Miju Kang$^2$ }
\affil{$^1$ School of Space Research, Kyung Hee University, Yongin-Si, Gyeonggi-Do 446-701, Republic of Korea \\
	   $^2$ Korea Astronomy and Space Science Institute, 776 Daedeokdaero, Yuseong, Daejeon 305-348, Korea\\      
	   $^3$ Square Kilometre Array Organisation, Jodrell Bank Observatory, Lower Withington, Cheshre SK11 9DL, United Kingdom \\
       $^4$ Department of Astronomy, University of Texas at Austin, 2515 Speedway, Stop C1400, Austin, TX 78712-1205, USA \\
       $^5$ National Research Council Canada, Herzberg Institute of Astrophysics, Victoria, BC, Canada \\
       %       $^6$ Institute for Astronomy, University of Hawaii at Manoa, Honolulu, HI 96822, USA  \\
       $^6$ Universidad Diego Portales, Facultad de Ingeniera, Av. Ej\'{e}rcito 441, Santiago, Chile \\
       $^7$ Harvard-Smithsonian Center for Astrophysics, 60 Garden Street, Cambridge, MA 02138, USA }

\begin{abstract}
We present a multi-wavelength observational study of a low-mass star-forming region, L1251-C, with observational results at wavelengths from the near-infrared to the millimeter. 
\textit{Spitzer Space Telescope} observations confirmed that IRAS 22343+7501 is a small group of protostellar objects.
The extended emission to east--west direction with its intensity peak at the center of L1251A has been detected at 350 and 850 \m\ with the CSO and JCMT telescopes, tracing dense envelope materials around L1251A.
The single-dish data from the KVN and TRAO telescopes show inconsistencies between the intensity peaks of several molecular line emission and that of the continuum emission, suggesting complex distributions of molecular abundances around L1251A.
The SMA interferometer data, however, show intensity peaks of \tcot\ and \thcot\ located at the position of IRS 1, which is both the brightest source in IRAC image and the weakest source in the 1.3 mm dust continuum map.
IRS 1 is the strongest candidate for the driving source of the newly detected compact \tcot\ outflow.
Over the whole region (14$\arcmin\times$14$\arcmin$) of L125l-C, 3 Class I and 16 Class II sources have been detected, including three YSOs in L1251A.  
A comparison with the average projected distance among 19 YSOs in L1251-C and that among 3 YSOs in L1251A suggests L1251-C is an example of low-mass cluster formation, where protostellar objects are forming in a small group.

\end{abstract}


\section{INTRODUCTION}

Many studies of low-mass star formation have been done in details.
Star formation mostly occurs in groups and clusters where the number density of young stellar objects (YSOs) at least $\sim$3 pc$^{-2}$ \citep{lada03}. 
The \textit{Spitzer} legacy project ``From Molecular Cores to Planets Forming Disks'' \citep[c2d;][]{evans03,evans09} found that most sites of low-mass star formation within 400 pc have significant populations of protostars forming in small groups of multiple objects (e.g., $<$10 members) located in close proximity.

L1251 is a star-forming region located in Cepheus at a distance of 300 $\pm$ 50 pc \citep{kun93} from the Sun. 
There are five dense C$^{18}$O cores A to E (N(C$^{18}$O) $\sim$ 10$^{15}$ cm$^{-2}$) in L1251, which have been assigned A to E with Right Ascension \citep{sato94}.
Two outflows, an extended \tco\ outflow and a well-collimated compact \tco\ outflow were previously known in cores C and E, respectively \citep{sato89}.
These two outflows are named L1251A and L1251B by \citet{hodapp94}.
Core C is the second densest C$^{18}$O core among five cores in L1251 \citep{sato94} where the strongest extended \tco\ outflow (L1251A) is located \citep{sato89}.
The dense region around IRAS 22376+7455, which is the driving source of the well-collimated compact outflow (L1251B) in core E has been studied in detail \citep{lee06,lee07}.  
Recently, another well-collimated outflow in core A has been revealed \citep{lee10}.
In this paper, we use ``L1251-C" to present the extended region observed in C$^{18}$O 1-0 and it is the same region with ``L1251C" in the \textit{Spitzer} data archive.
On the other hand, ``L1251A" is used to refer the densest region around IRAS 22343+7501, which is believed to be the driving source of the extended \tcot\ outflow. 
Our \textit{Spitzer} observations revealed that L1251A is a multi-protostellar system consisting of three sources.

Earlier studies already showed that IRAS 22343+7501 is composed of multiple protostellar objects.  
\citet{rosvick95} found that there are five near-infrared (NIR) sources labeled A--E around L1251A.
\citet{meehan98}, \citet{beltran01}, and \citet{reipurth04} presented also multiple radio continuum sources, detected with Very Long Array (VLA), in L1251A.
Two VLA radio sources, VLA 6 and VLA 7 \citep{beltran01,reipurth04} are associated with the NIR sources, D and A of  \citet{rosvick95}, respectively.
\citet{kun09} presented optical spectra and photometry for sources A and B of \citet{rosvick95} and classified both sources as classical T Tauri stars.

NH$_{3}$ cores were also detected around L1251A \citep{anglada97,toth96}.
\citet{nikolic03} presented the distributions of several molecular transitions compared with the NH$_{3}$ distribution of \citet{toth96} and showed outflow-like structures, which are possibly driven by VLA 6 and VLA 7, in \hcop.
\water\ masers were also detected \citep{toth94,wilking94,xiang95,claussen96} in this region.

\citet{schwartz88} discovered a bipolar \tco\ outflow in L1251A.
This \tco\ bipolar outflow shows asymmetric blue and red lobes in morphology and intensity.
\citet{sato89} also detected an extended \tco\ outflow in L1251A with symmetric structure.
Interestingly, the \tco\ outflow from \citet{sato89} has a northeast-southwest (NE--SW) axis, which is perpendicular to the northwest--southeast (NW--SE) axis found by \citet{schwartz88}.
An optical jet was found by \citet{balazs92} near L1251A, supporting L1251A as the origin of the extended \tco\ outflow.

In this paper, we analyse the properties of young stellar objects (YSOs) based on the c2d survey data and newly obtained molecular line data to study the environment of star formation in L1251-C, especially around L1251A.
In this study we use the ($\alpha$,$\delta$) coordinates of IRAS 22343+7501 as (22$^h$35$^m$23$^s$.4, +75{\degr}17{\arcmin}07{\arcsec}) in the J2000.0 epoch.

Our observations of L1251-C are summarised in section 2.
Results and analysis of the data are described in section 3 and section 4, respectively.
 In section 5, we discuss star formation in small groups and outflows in L1251A.
Finally, a summary of this paper is given in section 6.

\section{OBSERVATIONS AND DATA}

\subsection{\textit{ Spitzer Space Telescope} Data}

In the \textit{Spitzer} legacy program c2d, L1251-C was observed at 3.6, 4.5, 5.8, and 8.0 \m\ with the Infrared Array Camera \citep[IRAC;][]{fazio04} on 2004 October 18 and also observed at 24 and 70 \m\ with the Multi-band Imaging Photometer for \textit{Spitzer} \citep[MIPS;][]{rieke04} on 2004 September 24. 
The High-Dynamic Range (HDR) mode observations were obtained to get more accurate intensity levels of saturated sources.
In the HDR mode, a long-integration frame of 10.4 second and a short-integration frame of 0.4 second were taken at each position.
The description of the data processing details can be found in the c2d final delivery documentation \citep[see][]{evans07}\footnote{see http://irsa.ipac.caltech.edu/data/SPITZER/docs/spitzermission/observingprograms/legacy/}.
In this paper, we use the c2d photometric data catalog of YSOs including Two Micron All Sky Survey (2MASS) photometric data.
For the saturated sources, however, we use the photometry obtained with the HDR mode data.

\subsection{Radio Single-dish Observations and Data}

\subsubsection{Korean VLBI Network Observations} 

 Molecular line observations of L1251-C were carried out with the Korean VLBI Network (KVN) 21 m radio telescope at the Yonsei site, Seoul from 2012 October to 2013 March and at the Ulsan site, Ulsan in 2013 April. 
The central position of mapping observations was the position of IRAS 22343+7501. 
We observed the \hcop , \htcop , \nhp , and \hcn\ lines to map the cloud structure.
We also observed the \water\  6$_{16}$-5$_{23}$ maser and the \htco\  transition simultaneously.
The KVN radio telescopes were designed for simultaneous observations in the 22, 43, 86, and 129 GHz bands (Han et al. 2008). 
The pointing was checked every 2 hours using R Cas, a strong SiO maser source near the target source.

The system temperatures during the observations ranged from 60 K to 250 K (at 22 GHz), from 100 K to 250 K (at 86 GHz), and from 150 K to 400 K (at 129 GHz), depending on weather conditions and elevations. 
We used digital spectrometer in its 32 MHz mode, yielding velocity resolutions of 0.21 km s$^{-1}$ (4096 channels at 22 GHz), 0.21 km s$^{-1}$ (4096 channels at 86 GHz), and 0.27 km s$^{-1}$ (4096 channels at 129 GHz). 
The 1 $\sigma$ rms noise levels are presented in Table 1.
The data were calibrated using the chopper wheel method to get an antenna temperature $T^{*}_{A}$.

The beam Full Width at Half Maximum (FWHM) is $\sim$30 \arcsec\ at 86 GHz. 
The maps were sampled every a half of beam FWHM (15{\arcsec}).
We used the CLASS\footnote{see http://www.iram.fr/IRAMFR/GILDAS/doc/html/class-html/class.html.} reduction package from Institue de Radioastronomi Millim\'{e}trique (IRAM) for the data reduction. 
A summary of observations is presented in Table 1.

\subsubsection{Taeduk Radio Astronomy Observatory Observations}

Observations of \tco\ were carried out in 2013 February with the 14 m radio telescope of Taeduk Radio Astronomy Observatory (TRAO), equipped with 9 receivers arranged in 3 $\times$ 3 pattern with 50 $\arcsec$ spacing. 
The central position of target was the same as in the KVN observations.  
The system temperatures during observations ranged from 950 K to 1100 K depending on the weather conditions and source elevations.
The velocity resolution is 0.327 \kms\ and the averaged 1 $\sigma$ rms noise level is about 0.37 K.

The beam FWHM and beam efficiency are 50 \arcsec\ and 0.48 at 115 GHz, respectively.
We mapped $\thicksim$275$\arcsec\times$ 275$\arcsec$ region around L1251A, sampling every a half of beam FWHM (25{\arcsec}).
We also used CLASS reduction package from IRAM for data reduction.
A summary of observations is also presented in Table 1.

\subsubsection{Data from Other Observations}

The 350 \m\ continuum emission map \citep{wu07} of L1251-C was obtained using the SHARC-II camera on the Caltech Submillimeter Observatory (CSO).

The 850 \m\ continuum emission map was reprocessed by \citet{di08} with the data obtained from the James Clerk Maxwell Telescope (JCMT) Submillimeter Common User Bolometer Array (SCUBA). 
The data used in this work are from the public archive maintained by the Canadian Astronomical Data Centre\footnote{The Canadian Astronomical Data Centre is operated by the Dominion Astrophysical Observatory for the National Research Council of Canada.} (CADC).

\subsection{Radio Interferometer Observations}

\subsubsection{Submillimeter Array Observations}

The Submillimeter array \citep[SMA;][]{ho04} observations of L1251A were carried out on 2007 October 17 in the compact configuration with six antennas ($\theta_{\textrm{FWHM}} \sim$3 \arcsec). 
The quasar J2005+778 and J0102+584 were observed for the gain calibration.
Titan was observed for the flux calibration, and the quasar 3C 454.3 was observed for the pass-band calibration.
Natural weighting was used without any tapering to produce the data cubes.
The 230 GHz receiver was tuned to observe \tcot\ and \thcot\ in separate correlator windows of 2 GHz bandwidths.
Channel widths for \tcot\ and \thcot\ are 1.06 \kms\ and 0.28 \kms, respectively.
The rms noise levels of \tcot\ and \thcot\ are presented in Table 2.
The remaining chunk of 24 windows were assigned to observe the 1.3 mm (225.404 GHz) continuum emission.
The continuum image was processed with appropriate primary beam correction (the primary beam is $\sim$55 \arcsec\ at this frequency).
The 1 $\sigma$ rms level of 1.3 mm continuum map is 4.5 mJy beam$^{-1}$.
The IDL-based MIR software package\footnote{See http://cfa-www.harvard.edu/~cqi/mircook.html.} adapted for SMA was used for calibration and the MIRIAD \citep{sault95} was used for imaging and analysing.  
A summary of observations is shown in Table 2.

\section{RESULTS}

\subsection{Infrared Images}

\subsubsection{Central Triple System: L1251A}

IRAC 3.6, 4.5, 5.8 \m\ and MIPS 24 \m\ images of L1251A are presented in Figure 1.
Three IR sources, IRS 1, IRS 2, and IRS 3, were identified with the HDR mode observations in the IRAC band images.
On the other hand, they are unresolved in the MIPS band image (Figure 1d).
Multi components of L1251A at NIR were detected for the first time by \citet{rosvick95}, where five NIR sources were identified as labeled A--E near IRAS 22343+7501.
Unfortunately the coordinates for the five sources from \citet{rosvick95} are not sufficiently precise to associate them with the \textit{Spitzer} sources.
Comparisons of the relative positions show that \textit{Spitzer} sources IRS 1, IRS 2, and IRS 3 correspond to source D, source 4/B/C, and source 1/A
of \citet{rosvick95}, respectively.
IRS 1 is the brightest source among three components in the IRAC band images except in the IRAC 1 band (3.6 \m), while IRS 2 is the faintest one. 

\subsubsection{YSO Candidates in L1251-C}

 Figure 2 shows the IRAC three-color composite image of the 3.6, 4.5, and 8.0 \m\ bands.   
For identification of YSO candidates, \citet{evans07} excluded possible extra-galactic background sources based on the criteria derived with the color-magnitude diagrams of the reprocessed \textit{Spitzer} Wide-area Infrared Extragalactic Survey \citep[SWIRE;][]{lonsdale03} data.
According to the c2d YSOc catalog, 16 YSO candidates were identified in this region.
Three more sources were identified by HDR mode observations (IRS 1, IRS 2, and IRS 3).
Includingly added the three sources, 19 YSO candidates were identified in total, and Figure 2 shows our identification numbers for the sources around L1251-C on IRAC 4.5 \m\ image.
Labels 1,2, and 3 represent IRS 1, IRS 2, and IRS 3, respectively, and labels from 4 to 19 are assigned along with Right Ascension.
Coordinates and Spitzer flux densities of the YSO candidates identified in the \textit{Spitzer} images are presented in Table 3.

\subsection{Continuum Images, Molecular Line Spectra, and Maps}

\subsubsection{Radio Single-dish Continuum Emission}

Figure 3 shows the continuum images observed with CSO SHARC II at 350 \m\ and JCMT SCUBA at 850 \m.
Dust continuum condensations associated with L1251A have been detected in both wavelengths \citep{young06,wu07}.
With the effective beam FWHMs of $\sim$9 $\arcsec$ and $\sim$22.9 $\arcsec$, respectively, YSOs in L1251A were not resolved.
The peak positions of dust continuum emission in both wavelengths are located at the center of the three IR sources, suggesting that the dust continuum emission traces the dense core covering L1251A with peak intensities of 5.4 and 1.02 Jy beam$^{-1}$ at 350 and 850 \m, respectively.  
Dust continuum condensations in both wavelengths are seen elongated in the east--west (E--W) direction from L1251A.
The 850 \m\ dust continuum emission is likely optically thin and traces the dense and cold material within the molecular core.
Therefore, the total (gas+dust) mass of L1251A, $M_{\textrm{total}}$, was calculated with the 850 \m\ integrated flux, $F_{\nu}$, using 
\begin{equation}
M_{\textrm{total}} = \frac{F_{\nu}{d}^{2}}{{\kappa}_{\nu}B_{\nu}(T_{\textrm{dust}})},
\end{equation}
and assuming that all dust-continuum emission at 850 \m\ arises from dust and is optically thin.
Here, $d$ is the distance \citep[300 $\pm$ 50 pc;][]{kun93}, and $B_{\nu}$ is the Planck function for a blackbody of dust temperature $T_{\textrm{dust}}$. 
To estimate mass, we assumed $T_{\textrm{dust}} = 20$ K and adopted the mass opacity coefficient, ${\kappa}_{\lambda} = 0.1(250 \m / \lambda)^{\beta}$ cm$^{2}$ g$^{-1}$ \citep{hildebrand83} with $\beta =$ 1.8 \citep{hatchell13}.
This mass opacity assumes a gas-to-dust ratio of 100, and the estimated mass represents the total (gas+dust) mass of the  molecular core.
Since 350 \m\ is not in the Rayleigh-Jeans limit for this temperature regime, so we derived total mass only with the 850 \m\ flux.
The derived total mass is 4.98 $M_{\sun}$ with the integrated flux density, 6.18 Jy.
The aperture size used for the flux density is 60 \arcsec.

\subsubsection{Radio Single-dish Molecular Line Emission}

Figure 4 presents the integrated intensity maps of the \hcop, \htcop, \nhp, and \hcn, which all trace dense gas, made from the KVN single dish observations.
We integrated over the full widths of \hcop\ and \htcop, and all components of hyperfine structures were integrated for the integrated intensity maps of \nhp\ and \hcn.
The molecular line integrated intensity extends in the E--W direction in a manner similar to the distribution of dust continuum emission except \hcop, which may trace both infall and outflow (Figure 3).
The integrated intensity peaks of these molecular lines are offset from the continuum peak position of L1251A, and relatively optically thin lines such as \htcop\ and \nhp\ show bigger offsets.
The integrated intensity peak of \htcop\ is $\sim$30 $\arcsec$ to the south of L1251A (Figure 4b) and is coincident with the position of IRS 14. 
The integrated intensity peak of \nhp\ is $\sim$20 $\arcsec$ to the southwest of L1251A (Figure 4c).      
The integrated intensity peaks of relatively optically thick molecular emission such as \hcop\ and \hcn\ slightly shifted to southwest. 
The disagreements of positions between the intensity peaks of single dish dust continuum emission and the molecular line integrated intensity peaks might indicate that the distribution of molecular gas around L1251A is complicated because of 4 sources existing within about 25 $\arcsec$ (i.e., 0.04 pc). 
For example, heating by these sources can evaporate CO from the dust grain surfaces, which is a mother molecule of HCO$^{+}$ but a destroyer of N$_{2}$H$^{+}$.

Figure 5 shows the line profiles of the above molecular lines at the center position (IRS 1), as well as that of \htco.
The spectrum of \hcop\ toward L1251A shows a strong self-absorption dip with a classic blue asymmetric line profile, which has been believed as an infall indicator \citep{zhou93}.
These dips imply true self-absorption when the velocity of the line peak of optically thin tracers such as \htcop, \htco, and \nhp\ coincides with the velocity of dip; although the centroid velocities of optically thin tracers such as \htcop\ and \htco\ are slightly blueshifted, they peak at the velocity of the self-absorption dip.
Gaussian fit FWHMs line widths of \htcop\ and \htco\ are 1.3 and 1.9 km s$^{-1}$, respectively.
From simultaneous hyperfine structure fitting, the FWHMs of  \nhp\ and \hcn\ are 0.6 and 1.04 km s$^{-1}$, respectively.
Therefore, \nhp\ seems to trace the most quiescent gas component.
In addition, the profile of $F=$ 2--1 hyperfine component of \hcn\ line, which also has been believed as an infall tracer, is similar to that of \hcop.

The line profile of \hcop\ is superimposed by two (broad and narrow) components and the narrow component shows the self-absorption feature.
The broad component ($v_{\textrm{cen}} \sim$ --4.8 km s$^{-1}$, $\Delta v \sim 2.9$ \kms) is seen in all positions across the observed map while the narrow component ($v_{\textrm{cen}} \sim$ --5.3 km s$^{-1}$, $\Delta v \sim 0.8$ \kms) is detected just around L1251A. 
The integrated intensity map of the broad component (not shown) shows a similar distribution to the \hcop\ outflow (see section 4.4 below), suggesting the origin of the broad component may be the outflows from L1251A.
The integrated intensity map of the \hcop\ narrow component and that of the isolated component of \nhp\ show similar distributions, although the integrated intensity peaks of the two lines are not coincident (Figure 6).
The similarity suggests that the narrow component of \hcop\ may trace the dense and relatively quiescent molecular gas around L1251A like the other molecular lines (except for  \tco).

\subsubsection{Radio Single-dish Water Maser Emission}

\water\ maser emission at 22 GHz was monitored toward L1251A to find variations of the systemic velocity and intensity in time. 
In Figure 7, multi-epoch results are presented.
Maser components of different velocity or intensity were seen at different epochs.
Five redshifted maser components can be identified at --2.4, 0, 2, 3.2, and 153 km s$^{-1}$.
The --2.4 km s$^{-1}$ component was presented in 2013 February, slightly weakened in 2013 March and become very strong in 2013 May. 
The  0 km s$^{-1}$ component was first detected in 2012 December and became stronger in 2013 January but weakened in 2013 May. 
The 3.2 km s$^{-1}$ velocity component and the very high velocity component 153 km s$^{-1}$ were also detected in 2013 February (Figure 8).   
The integrated intensity peak of the \water\ maser emission (not shown) located in L1251A but the beam FWHM is not small enough to identify the specific location of the water maser source.

The isotropic luminosity of an \water\ maser can be estimated by the integrated intensities of observed emission using the following equation \citep{bae11}:
\begin{equation}
L_{\textrm{H$_{2}$O}} = 4{\pi}d^{2}\frac{v}{c}\int{F_{\nu}dv}.
\end{equation} 
Here, $d$ is the distance \citep[300 $\pm$ 50 pc;][]{kun93}, and $\nu$ is the observed frequency.
When multiple lines are present, the sum of all the individual lines are used to calculate the luminosity.
The derived \water\ maser luminosities are presented in Table 4.

\subsubsection{Radio Interferometer Continuum Emission}

The 1.3 mm continuum image obtained with the SMA toward L1251A is presented in Figure 9.
 Three individual condensations were detected.
For the 1.3 mm continuum sources, we will follow the nomenclature of \citet{reipurth04}. 
Although the peak positions of submillimeter sources are not exactly coincide with the positions of IR sources, VLA 6 and VLA 7 are the same sources as IRS 1 and IRS 3, respectively.
However, VLA 10 is apart from IRS 2 by about 2 \arcsec.
The positions of IRS 2 and the 2MASS source are consistent.
Therefore, VLA 10 and IRS 2 are tracing two different sources.
The dust continuum emission related to VLA 6 and VLA 7 was first detected by \citet[their VLA A and VLA B]{meehan98} at 3.6 cm and next by \citet{beltran01} at 3.6 cm and 6 cm.
\citet{reipurth04}, however, first detected the continuum condensation associated with VLA 10 at 3.6 cm, and they mentioned that VLA 10 is a radio jet source at a position angle of about 53 $\degr$, indicating that VLA 10 is another protostellar member in L1251A although no corresponding IR source has been detected.
\citet{reipurth04} also mentioned that VLA 7 is a radio jet source at a position angle of about 126 $\degr$.

Although VLA 6 and VLA 10 are connected at low intensity levels, three sources are resolved in our 1.3 mm continuum map.
To estimate their masses, we used $T_{\textrm{dust}} = 30$ K \citep{jorgensen07} and adopted the dust mass opacity coefficient, ${\kappa}_{\lambda} = 0.02(1~mm / \lambda)^{\beta}$ cm$^{2}$ g$^{-1}$ \citep{beckwith91}, which is normalized for flat sources ($\beta =$ 1).
The masses of VLA 6, VLA 7, and VLA 10 are presented in Table 5.
The total mass derived from the SCUBA 850 \m\ continuum map is roughly 55 times larger than the sum of masses derived from the SMA 1.3 mm continuum sources because the single dish flux includes extended emission around all four sources.

\subsubsection{Radio Interferometer Molecular Line Emission}

Line profiles of \tcot\ and \thcot, observed with the SMA at the position of each submillimeter source are presented in Figure 10.
Channel widths for \tcot\ and \thcot\ are 1.09 \kms\ and 0.23 \kms, respectively.
Among the three submillimeter sources, the strongest peak brightness temperature of \tcot\ (24.75 K) is seen at VLA 6, while the peak brightness temperatures of \thcot\ ($\sim$7 K) are similar toward all three sources. 
The centroid velocity of \thcot\ is shifted from red to blue in the order of VLA 6, VLA 10, and VLA 7. 
There are channels without emission in \thcot\ at \vlsr\ = --4.94 and --3.84 \kms.
The two dips of \thcot\ cannot be resolved with the $\sim$1 \kms\ velocity resolution of \tcot\ data.
Moreover, \tcot\ emission is missing in the velocity range of \vlsr\ = --6 to --4 \kms.
At these velocities, most emission from the larger-scale structure in that velocity range has been likely resolved out.

In Figure 11, the velocity channel map of the \tcot\ line is presented.
The \tcot\ emission is in the velocity range of --14.9 to 5.2 \kms.
In the channel map, the \tcot\ emission is missing at the channels of --5.4 and --4.3 \kms. 
Moreover, the strong negative contours are shown near bright emission, caused by the missing flux effect at --3.3 and --2.2 \kms.
A blueshifted component related to VLA 6 or VLA 10 is seen extending to the southwest of L1251A. 
A strong redshifted component is located at the position of VLA 6, extending in the northeast direction.
Though it is hard to distinguish apparent bipolar outflow structures, we infer that the blue- and redshifted components are from outflowing gas, because the velocity offsets from the systemic velocity are $\geq 2.5$ \kms.
We will discuss more about outflows in section 4.4 below.
Figure 12 shows the \tco\ integrated intensity across L1251A, overlaid onto the 1.3 mm continuum emission map for reference. 
The integrated intensity map of \tco\ shows E--W elongation, and the intensity peak is located at the position of VLA 6.

Figure 13 shows the velocity channel map of the \thcot\ line.
Every four channels are integrated for \thcot\ to compare with the \tcot\ channel map.
The redshifted emission at the position of VLA 6 shown at --3.1 \kms, while the blueshifted emission at the position of VLA 10 and VLA 7 is shown at --5.4 and --6.5 \kms, respectively.
The blueshifted emission at --5.4 \kms\ is slightly extended to the southwest and the blueshifted emission at --6.5 \kms\ is extended to the northeast.
At --5.4 \kms, there is another blob to the east of VLA 10, which can be discerned in the integrated intensity map of \thcot. 
In Figure 14, the integrated intensity map of \thcot\ across L1251A is presented and also overlaid onto 1.3 mm continuum emission map for reference.
Each continuum source seems to be associated with a \thcot\ integrated intensity peak although those peaks are not completely distinct.
The whole emission structure extended along the NE--SW direction.

Most emission of blue- and redshifted components of \thcot\ is seen in the velocity range where the \tcot\ emission appears to have been resolved out.
In addition, the blue- and redshifted components of \thcot\ have low velocities (within \vlsr\ $\pm$ 3 \kms) compared to that of \tcot\ (within \vlsr\ $\pm$ 9 \kms).
Therefore, the features of \thcot\ seem to trace the innermost part of envelopes around each submillimeter source rather than outflows.

\section{ANALYSIS}

\subsection{YSO Classification}

Except IRS 1, IRS 2, and IRS 3 identified by the HDR observations, the IR spectral index $\alpha$ of 16 YSO candidates are listed in the c2d YSOc catalog.
The $\alpha$ for IRS 1, IRS 2, and IRS 3 are calculated in this work.
All photometries between 2 \m\ and 24 \m\ in the c2d catalog are used to determine the IR spectral index $\alpha$  \citep{evans09}:
\begin{equation}
\alpha = \frac{d\ \textrm{log}(\lambda S(\lambda))}{d\ \textrm{log}(\lambda)}.
\end{equation}
Here, $\lambda$ is the wavelength and $S(\lambda)$ is the flux density at that wavelength.
Following the criteria of \citet{evans09} (i.e., Class I for 0.3 $\leqslant \alpha$, Flat --0.3 $\leqslant \alpha <$ 0.3, Class II for --1.6 $\leqslant \alpha <$ --0.3, and Class III for $\alpha <$ --1.6), IRS 1, 11, and 16 are Class I candidates, IRS 3, 7, and 17 is Flat class candidates, and all others are Class II candidates.
Table 6 lists the IR spectral index determined for each source.
We calculated the IR spectral indices of the newly added YSOs, also listed in Table 6. 
We also applied two other criteria of classification (the color-color diagram and $T_{\textrm{bol}}$) to all 19 YSOs. 
Photometric data of all YSOs are presented in Table 3 from the \textit{Spitzer} observations.
Figure 15 shows a [3.6] - [4.5] versus [5.6] - [8.0] color-color diagram used to classify the evolutionary stages of YSO candidates, including all identified sources in Figure 2.
We adopted the same criteria used in \citet{lee06}, which is basically based on the models of \citet{allen04}.
Two sources, IRS 1 and 14, are identified as Class 0/I candidates in color-color diagram.
IRS 11 and 17 are located at the color-color boundaries between Class 0/I and Class II.
The IR spectral indices of these sources, however, indicate that IRS 11 is in Class I and IRS 17 is in Flat class.
IRS 4 is located close to the boundary of Class III criteria, but its IR spectral index of IRS 4 is in the range for Class II.

In Table 6, derived bolometric luminosities ($L_{\textrm{bol}}$) and bolometric temperatures ($T_{\textrm{bol}}$) are listed. 
If the classification by \citet{myers93} (i.e., Class 0 for $T_{\textrm{bol}} < 70$ K, Class I for 70 K $\leqslant T_{\textrm{bol}} \leqslant$ 650 K, and Class II for 650 K $< T_{\textrm{bol}} \leqslant$ 2800 K) is used, IRS 1, 11, and 16 are identified as Class I, and the other 16 YSOs are identified as Class II.
The evolutionary stages of most sources identified by $T_{\textrm{bol}}$ agrees with those by the IRAC color-color diagram.
Note, however, that IRS 14 is identified as an embedded source by the color-color diagram but as Class II by $T_{\textrm{bol}}$ and $\alpha$.
It is more likely in Class II, but since it is spatially located at the \htcop\ emission peak it may be Class 0/I instead.
Furthermore, IRS 11, which is located at the boundaries of Class 0/I and Class II in the IRAC color-color diagram, is identified as Class I by $T_{\textrm{bol}}$ (IRS 11 has a very low luminosity).
In addition, IRS 4 is identified as Class II source by $T_{\textrm{bol}}$.
Finally, IRS 16 is identified as a Class I source with the lowest $T_{\textrm{bol}}$ among 19 YSO candidates.

IRS 11 and IRS 16 are identified as Class I YSOs with luminosities smaller than 0.1 L$_{\sun}$, suggesting these sources are candidates of Very Low Luminosity Objects \citep[VeLLOs,~e.q.,][]{young04,bourke05,dunham06}.  
However, it cannot be ruled out that these two sources may be background galactic sources. 
\citet{evans07} defined the probability of being background galactic source, Log(Prob$\_$Galc), and galaxy candidates have Log(Prob$\_$Galc) $\geqq$ --1.47.  For IRS 11 and IRS 16, the probability is --2.17 and --1.58, respectively.

Spectral energy distributions (SEDs) for the YSOs in L1251-C are presented in Figure 16.
The SEDs include 2MASS, IRAC, and MIPS bands data.
IRS 1 is the brightest source at near- and mid-infrared wavelengths except in IRAC 1 band (3.6 \m).

\subsection{L1251A and the Common Envelope}

Including the radio continuum source VLA 10, the number of members in L1251A is four.
The mean projected separation between each companion is about 1220 AU. 
\citet{looney00} classified multiple systems by their morphologies into three groups: separate envelope system, common envelope system, and common disk system.
The typical separation of common envelope systems is from 150 AU to 3000 AU. 
The projected distances among members of this system suggest that L1251A is a common envelope system, having one primary core of gravitational concentration.

\subsection{Molecular Outflows}

In L1251-C, an extended bipolar \tco\ outflow driven by IRAS 22343+7501 was reported by \citet{sato89}.
An optical jet was also observed \citep{balazs92} with the same axis (NE--SW) as the extended \tco\ outflow. 
The outflow map of \citet{schwartz88}, however, shows a bipolar structure with an unclear outflow axis in a small region around L1251A.

For this paper, the velocity ranges for blue- and redshifted outflow wings were defined as follows.
The outer velocities of the ranges are determined for each position where the emission is greater than the respective 1$\sigma$ rms noise level. 
Spectra at outflow-free positions, chosen by eye, are used to define the inner velocities of the ranges, as being equal to the outermost velocities.

The averaged outflow-free spectra of \tco\ for the blue and red component are presented in Figure 17a. 
The spectra at the peak positions of the blue- and red-outflow components are also shown.
The outflow map of \tco\ made by integrating intensity over the determined outflow wings is presented in Figure 17b. 
The outflow wing emission is elongated mainly in the east--west (E--W) direction, and it also shows asymmetry in intensity and morphology.
For example, the red component of the \tco\ outflow is more broadly extended than the blue component.
Also, the blue peak is located at the east of L1251A, and the red peak is located at the north of that region.

In comparison, a \hcop\ outflow map shows blue and red components with significant asymmetry in intensity and size. 
The velocity ranges for blue- and redshifted outflow wings of \hcop\ are --9.7 to --5.9 \kms\ and --4.0 to 0.2 \kms, respectively.
Figure 18 shows the blue component is dominant. 
A poorly collimated outflow of \hcop\ is extended along the NW--SE direction, i.e., inconsistent with the E--W direction of the \tco\ outflow.
The position angle of the \hcop\ outflow axis, however, does correspond to that of the \hcop\ outflow-like structure seen by \citet{nikolic03}.

Figure 19 shows the outflow of \tcot\ detected with the SMA. 
The velocity ranges for blue- and redshifted outflow wings of \tcot\ are --20 to --7 \kms\ and from --2 to 10 \kms, respectively.
Then the axis of this small scale outflow is not consistent with either the \tco\ outflow or the \hcop\ outflow. 
The blue- and redshifted emission extends along the NE--SW direction and is very compact.
IRS 1 located at the position of the red peak of the \tcot\ outflow and IRS 1 is the closest to the center of the blue and red components.
In addition, IRS 1 is in the earliest evolutionary stage among three YSOs of L1251A detected in IRAC observations (see 5.1 below), suggesting IRS 1 may be the driving source of the compact \tcot\ outflow.
However, in spite of no corresponding IR source, a radio jet has been detected in VLA 10 and its propagation direction is consistent with the compact outflow direction.
Therefore, higher-resolution observations are needed to determine the engine of this compact outflow.

Different molecular outflows on different scales show dissimilarities in axis, morphology, and strength (Figure 17, 18, and 19), suggesting that there are multiple outflows from L1251A.
The extended \tco\ outflow from \citet[][see~there~Figure~3]{sato89} has a direction completely opposite to that of the compact \tcot\ outflow presented here.
The \hcop\ and \tco\ outflows presented here may show a mixture of extended and compact outflows with poor collimation.

\section{Discussion}
\subsection{Star Formation in Small Groups}

In L1251-C, 19 YSOs are detected, including three that are identified in the HDR mode observations.
The submillimeter source, VLA 10, is excluded for this analysis, because VLA 10 was not detected by \textit{Spitzer} observations.
The fraction of identified sources as Class 0/I YSOs is 16 $\%$ and 84 $\%$ of sources are identified as Class II, based on $T_{\textrm{bol}}$ classification (see Table 6).
In L1251A, however, one is a Class 0/I source and two others are Class II sources. 
The whole field size of the MIPS observations of L1251-C is $\sim$1.44 pc$^{2}$, therefore the surface density of YSOs in L1251-C is $\sim$13.2 pc$^{-2}$. 
The surface density of YSOs in L1251A is larger than that of the whole L1251-C region by two orders of magnitude.
The area used for calculating the YSO surface density of L1251A is $\sim$0.08 pc$^{2}$, which covers its emission at submillimeter wavelengths (see Figure 3).

 We compare L1251-C with the other core in L1251 \citep{sato94} and also observed with \textit{Spitzer} \citep{lee06}.
\citet{lee06} shows that the surface density of YSOs in L1251-E is close to those in active star-forming regions.
L1251-E ($\sim$2.5 pc$^{2}$) has a larger area than L1251-C ($\sim$1.44 pc$^{2}$).
In L1251-E, 20 YSO candidates (4 Class 0/I, 14 Class II, and 3 unclassified YSOs) were detected in the MIPS 24 \m\ band. 
As a result, in L1251-E, the surface density of YSOs is $\sim$10 pc$^{-2}$, which is slightly lower than that of L1251-C by a factor of 1.3.
L1251B, the densest part in L1251-E, however, shows a larger YSO surface density than that of L1251A by a factor of 2.
The composition of YSOs in various evolutionary stages in L1251-E and L1251-C shows similar fractions.
For example, the fractions of Class 0/I candidates in L1251-C and L1251-E are about 16 $\%$ and 22 $\%$, respectively.
The fraction of Class 0/I sources in L1251B (50 $\%$), is slightly larger than that in L1251A ($\sim$34 $\%$). 
Overall, L1251-C has a similar star formation environment to L1251-E.

One Class 0/I YSO and two Class II YSOs are located in L1251A in the IRAC HDR images (see, e.g., Fig. 1).
The average projected distance among the 19 YSOs in L1251-C is about 136 $\arcsec$, while the average projected distance of three YSOs in L1251A is about 5.6 $\arcsec$, suggesting clustering.
The intensity peak of single-dish submillimeter continuum emission, which is coincident with all three YSOs of L1251A, is located between all three YSOs, suggesting that three YSOs at different evolutionary stages all reside in a singe dense core.
Interferometer submillimeter molecular line emission from L1251A observed with SMA, which shows more detailed view of the dense gas around the YSO group and resolves out the large r-scale structure, has their intensity peaks located at the position of IRS 1, showing IRS 1 is more deeply embedded than the other two YSOs.
Single-dish submillimeter molecular line emission from L1251A, however, has their integrated intensity peaks shifted to the south of L1251A, probably because of CO evaporation.

The interferometric continuum data at 1.3 mm are likely tracing disks around protostars, but the contamination from the inner envelope cannot be ruled out because the deconvolved sizes of submillimeter sources are similar or slightly larger than the beam FWHM (2.7 \arcsec $\times$ 2.5 \arcsec). 
The masses derived from the 1.3 mm continuum decrease in order of VLA 10, VLA 7, and VLA 6 (0.030, 0.023, and 0.020 M$_{\sun}$, respectively).
VLA 10 does not have a corresponding IR source, indicative of a deeply embedded source. 
Therefore, the 1.3 mm continuum emission of VLA 10 might trace its very dense inner envelope.

\subsection{Outflows in L1251A}

At least two outflows with different properties exist in L1251-C.
One is an extended \tco\ outflow \citep{sato89} and the other is a compact \tcot\ outflow (this paper). 
The extended outflow from \citet[][see~their~Figure~3]{sato89} shows a completely opposite outflow direction to that of the compact outflow from this work.  
The dynamical time of the extended outflow (1.7$\times10^{5}$ yr) is greater than that of the compact outflow (4.7$\times10^{3}$ yr) by two orders of magnitude.
Submillimeter continuum associated with radio jets (VLA 7 and VLA 10) is detected in L1251A.
VLA 7 has the NW--SE propagation direction and VLA 10 has the NE--SW propagation direction.   
Poorly collimated distribution of blue- and redshifted components of KVN \hcop\ outflow and TRAO \tco\ outflow may be also affected by these radio jets of VLA 7 and VLA 10.

The driving sources of each outflow, however, are hard to distinguish.
A strong candidate of the driving source of the compact outflow is IRS 1 (VLA 6) because it is an embedded Class I source, and the blue- and redshifted lobes are closely associated with the position of IRS 1 (see Figure 19). 
However, VLA 10 cannot be ruled out because its dust continuum emission seems to trace a very dense inner envelope (indicative of an extremely embedded source), and it is associated with a radio jet, whose propagation direction is consistent with that of the compact outflow \citep{reipurth04}.

\section{SUMMARY}

L1251-C was observed as part of the \textit{Spitzer} legacy project, c2d.
Also, various molecular emission lines associated with L1251A have been observed with the KVN 21 m radio telescope and TRAO 14 m radio telescope to understand its environment.
Single-dish continuum data from other observations are also presented.  
In addition, SMA interferometric continuum and line data of L1251A were obtained at high angular resolution.
With these data, we find:

1. 19 YSO candidates are associated with L1251-C, including three additionally identified in the HDR mode \textit{Spitzer} IRAC bands: IRS 1, IRS 2, and IRS 3.

2. Single-dish continuum emission at 350 and 850 \m\ is extended to the E--W and may trace an envelope containing all three sources. 
Individual condensations associated with each IR source are not resolved in the dust continuum map at 350 and 850 \m.
L1251A was mapped at KVN in \hcop, \htcop, \nhp, and \hcn.
The integrated intensity peaks of all molecular lines observed at KVN are offset from the peak positions of the continuum maps at 350 and 850 \m.
Notably, the integrated intensity maxima of \htcop\ and \nhp, which are optically thin, are offset by tens of arcseconds from L1251A to the south or southwest.
The line profile of a well known outflow and infall tracer \hcop\ line are composed of two velocity (broad and narrow) components and show a self-absorption feature.
Though this feature may trace infall, the broad component is a sign of outflow and the narrow component may trace dense envelope material (like other molecular lines described here).
Several velocity components of the \water\ 22 GHz maser have been detected during multi epoch observations.

3. Three dust continuum condensations (VLA 6, VLA 7, and VLA 10) are detected by the SMA observations at 1.3 mm.
Two of three (VLA 6 and VLA 7) are detected at the positions of IR sources (IRS 1 and IRS3, respectively) in L1251A. 
The positions of IRS 2 and VLA 10 are apart by $\sim$2 \arcsec, suggesting they are different sources.
Finally, the total number of protostars in L1251A is four by adding the most embedded source of VLA 10, which does not have a corresponding IR source.
The SMA integrated intensity maps of \tcot\ and \thcot\ show complicated distributions although a compact outflow is clearly detected in \tcot\ in the NE--SW direction.

4. In L1251-C, IRS 1, IRS 11, and IRS 16 are identified as Class I, and the other 16 YSOs are identified as Class II.
The surface density of L1251-C ($\sim$1.44 pc$^{2}$) is similar to that of L1251-E ($\sim$2.5 pc$^{2}$).
The fraction of Class 0/I sources (16 $\%$) in L1251-C is also similar to that seen in L1251-E (22 $\%$).

5. Molecular outflows on different scales have been mapped in \hcop, \tco, and \tcot. 
These outflows lack coincidence in axis, morphology, and strength, suggesting there are two outflows from two sources. 
One is a extended \tco\ outflow from \citet{sato89} and the other is a compact SMA \tcot\ outflow described here. 
KVN \hcop\ and TRAO \tco\ observations may show a mixture of extended and compact outflows.
IRS 1 (VLA 6) is likely the driving source of the compact outflow detected in higher-resolution SMA \tcot\ observations, but VLA 10 cannot be ruled out.
Therefore, interferometric observations with still higher spatial resolution are necessary to identify better which source is driving the compact outflow.

\clearpage

Support for this work, part of the \textit{Spitzer} Legacy Science Program, was provided by NASA through contracts 1224608 and 1230782 issued by the Jet Propulsion Laboratory, California Institute of Technology, under NASA contract 1407. 
This research was also supported by the Basic Science Research Program through the National Research Foundation of Korea (NRF) funded by the Ministry of Education of the Korean government (grant No. NRF-2012R1A1A2044689).
J.-E. L. is very grateful to the department of Astronomy, University of Texas at Austin for the hospitality
provided to her from August 2013 to July 2014.
This work was supported by the BK21 plus program through the National Research Foundation (NRF) funded by the Ministry of Education of Korea.
This work was also supported by the Korea Astronomy and Space Science Institute (KASI) grant funded by the Korea government (MEST).
This research was supported in part by the National Science Foundation under grant number 0708158, and by NASA under grant number NNX13AE54G (T.L.B.).

\begin{deluxetable}{lrccccl}
\tabletypesize{\footnotesize}
\tablecaption{Summary of Radio Single-Dish Observations}
\tablewidth{0pt}
\tablehead{	\colhead{} & 
			\colhead{$\nu$} &  
			\colhead{$\Delta$ V} & 
			\colhead{Beam FWHM} & 
			\colhead{$\eta_{mb}$} & 
			\colhead{1 $\sigma$ rms} &
			\colhead{Telescope} \\
			\colhead{Molecular line} & 
			\colhead{[GHz]} & 
			\colhead{[km s$^{-1}$]} & 
			\colhead{[arcsec]} &
			\colhead{} &
			\colhead{[K]} &
			\colhead{}}
		
\startdata
{\water\ 6$_{16}$-5$_{23}$} & 22.235080 & 0.21 & 120 & 0.38 & 0.08 & KVN (YS) \\
& & & & 0.42 & 0.09 & KVN (US) \\ 
{\htcop} & 86.754288 & 0.22 & 31.5 & 0.42 & 0.08 & KVN \\
{\hcn} & 88.631847 & 0.22 & 31.5 & 0.41 & 0.09 & KVN \\
{\hcop} & 89.188526 & 0.22 & 31.5 & 0.40 & 0.09 & KVN \\
{\nhp} & 93.176265 & 0.22 & 31.5 & 0.37 & 0.09 & KVN \\
{\htco} & 140.839520 & 0.27 & 23.4 & 0.30 & 0.11 & KVN \\
{\tco}  & 115.271202 & 0.21 & 50 & 0.48 & 0.37 & TRAO 
\enddata

\end{deluxetable}

\clearpage

\begin{deluxetable}{lrcccccc}

\tablecaption{Summary of SMA Observations}
\tabletypesize{\footnotesize}
\tablewidth{0pt}
\tablehead{	\colhead{} & 
			\colhead{$\nu$} &  
			\colhead{} & 
			\colhead{Number of} & 
			\colhead{$\Delta$ V} &
			\colhead{Beam FWHM} &	
			\colhead{P.A.} &
			\colhead{1 $\sigma$ rms}  \\
			\colhead{Molecular line} & 
			\colhead{[GHz]} & 
			\colhead{Side Band} & 
			\colhead{Channels} & 
			\colhead{[km s$^{-1}$]} &
			\colhead{[arcsec $\times$ arcsec]} &			
			\colhead{[deg]} &
			\colhead{[Jy beam$^{-1}$]}}
		
\startdata
{\tcot} & 230.5379700 & USB & 128 & 1.06 & 3.6 $\times$ 3.1 & -1.3 & 0.20  \\
{\thcot} & 220.3986765 & LSB & 512 & 0.28 & 3.7 $\times$ 2.3 & -5.8 & 0.37
 \enddata

\end{deluxetable}

\begin{deluxetable}{ccccccccc}
\tabletypesize{\tiny}
\tablewidth{0pt}
\rotate
\tablecaption{Infrared Fluxes of Young Stellar Object Candidates from the Spitzer Data }
\tablehead{ \colhead{} & 
			\colhead{RA} &
			\colhead{Dec} &       
			\colhead{IRAC 3.6 $\mu$m} &
			\colhead{IRAC 4.5 $\mu$m} &
			\colhead{IRAC 5.8 $\mu$m} &
			\colhead{IRAC 8.0 $\mu$m} & 
			\colhead{MIPS 24 $\mu$m} &
			\colhead{MIPS 70 $\mu$m} \\
			\colhead{Source} &
			\colhead{[$^{h}$ $^{m}$ $^{s}$]} &
			\colhead{[$\degr$ $\arcmin$ $\arcsec$]} &
			\colhead{[mJy]} &
			\colhead{[mJy]} &
			\colhead{[mJy]} &
			\colhead{[mJy]} &
			\colhead{[mJy]} &
			\colhead{[mJy]} }

\startdata 
1 & 22:35:23.28 & +75:17:07.73 & 293.00 (13.00) & 748.00 (13.00) & 1,550.00 (32.00)  & 2,510.00 (39.900) & \nodata & \nodata \\
2 & 22:35:24.57 & +75:17:03.61 & 183.00 (3.20) & 114.00 (3.60) & 111.00 (3.30) & 128.00 (1.30) & \nodata & \nodata \\
3 & 22:35:24.96 & +75:17:11.33 & 532.00 (22.00) & 610.00 (11.00) & 707.00 (14.00) & 890.00 (19.00) & \nodata & \nodata \\
4 & 22:34:12.00 & +75:18:09.36 & 183.00 (58.90) & 114.00 (14.60) & 111.00 (34.90) & 128.00 (31.40) & 1,440.00 (135.00) & 1,740.00 (162.00) \\
5 & 22:34:40.56 & +75:17:44.16 & 30.20 (1.51) & 28.40 (1.38) & 27.70 (1.36) & 35.60 (1.40) & 139.00 (12.90) & 257.00 (24.50) \\
6 & 22:34:48.24 & +75:13:35.04 & 8.97 (0.44) & 7.30 (0.35) & 6.68 (0.32) & 9.04 (0.43) & 17.80 (1.66) & 50.90 (5.83)  \\
7 & 22:35:00.72 & +75:15:36.36 & 22.00 (1.13) & 23.40 (1.13) & 25.70 (1.25) & 32.30 (1.52) & 102.00 (9.44) & 146.00 (14.50) \\
8 & 22:35:02.40 & +75:17:58.20 & 35.60 (1.83) & 43.10 (2.13) & 45.60 (2.23) & 47.60 (2.30) & 65.30 (6.06) & \nodata  \\
9 & 22:35:04.08 & +75:18:20.52 & 32.40 (1.64) & 32.00 (1.54) & 36.00 (1.74) & 40.80 (1.95) & 80.30 (7.44)  & 106.00 (10.70) \\
10 & 22:35:09.60 & +75:16:09.48 & 7.96 (0.40) & 8.40 (0.41) & 6.19 (0.32) & 5.13 (0.24) & 12.70 (1.19) & \nodata  \\
11 & 22:35:14.06 & +75:15:02.39 & 1.50 (0.08) & 1.69 (0.11) & 2.56 (0.12) & 20.00 (1.86) & 146.00 (14.30) & \nodata \\ 
12 & 22:35:16.56 & +75:18:46.80 & 75.90 (4.02) & 74.20 (3.81) & 73.00 (3.64)  & 85.90 (4.15) & 107.00 (9.96) & 58.50 (6.78) \\
13 & 22:35:25.44 & +75:17:56.04 & 102.00 (5.39) & 108.00 (5.38) & 121.00 (6.05) & 123.00 (5.97) & 75.50 (7.45) & \nodata \\
14 & 22:35:26.62 & +75:16:36.86 & 34.50 (1.78) & 53.20 (2.64) & 58.00 (2.81) & 54.10 (2.71) & 67.90 (8.19) & \nodata \\
15 & 22:35:27.12 & +75:18:01.80 & 37.20 (1.85) & 34.20 (1.66) & 33.00 (1.64) & 3.55 (1.70) & 52.90 (5.25) & \nodata \\
16 & 22:35:59.21 & +75:17:50.40 & 0.071 (0.007) & 0.078 (0.007) & 0.084 (0.024) & 0.118 (0.020) & 7.820 (1.760) & \nodata \\
17 & 22:36:05.52 & +75:18:32.04 & 78.20 (4.11) & 102.00 (5.30) & 145.00 (7.29) & 246.00 (12.20) & 534.00 (45.60) & 377. (36.60) \\
18 & 22:36:35.52 & +75:21:34.92 & 54.80 (2.72) & 54.90 (2.77) & 58.10 (2.81) & 72.60 (3.65) & 206.00 (19.10) & \nodata \\
19 & 22:36:59.04 & +75:21:20.52 & 2.87 (0.19) & 2.38 (0.14) & 2.11 (0.13) & 2.06 (0.11) & 2.56 (0.30) & \nodata
\enddata

\end{deluxetable}

\clearpage

\begin{deluxetable}{cc}
\tabletypesize{\footnotesize}
\tablecaption{The isotropic luminosity of H$_{2}$O maser}
\tablewidth{0pt}
\tablehead{ \colhead{Observing Date} &
		    \colhead{$L_\textrm{H$_{2}$O}$} \\
			\colhead{} &
			\colhead{[$L_\sun$]}}	
\startdata
2012 Dec & 9.46E--09  \\
2013 Jan &  5.55E--09  \\
2013 Feb & 7.41E--08  \\
2013 Mar &  9.59E--08  \\
2013 May &  2.01E--07 \\
\enddata

\end{deluxetable}

\begin{deluxetable}{lcccccc}
\rotate
\tabletypesize{\footnotesize}
\tablecaption{Parameters of the SMA 1.3 mm Dust Continuum Condensations Associated with L1251A  }
\tablewidth{0pt}
\tablehead{ \colhead{ID} &
		        \colhead{ RA } &
		        \colhead{ Dec } &
			\colhead{ Peak Flux} &
			\colhead{ Integrated Flux Density} &
			\colhead{ Source Size } &
			\colhead{ Mass } \\
			\colhead{} &
			\colhead{[$^{h}$ $^{m}$ $^{s}$]} &
			\colhead{[$\degr$ $\arcmin$ $\arcsec$]} &
			\colhead{[mJy beam$^{-1}$]} &
			\colhead{[mJy]} &
			\colhead{[$\arcsec$ x $\arcsec$] (P.A. $\degr$ )} &
			\colhead{[$M_{\sun}$}]}
			
\startdata 

VLA 6 (IRS 1) & 22:35:23.45 & +75:17:07.73 & 24.20 (0.001) & 35.23 & 3.9 $\times$ 2.5 (--0.9) & 0.024\\
VLA 7 (IRS 3) & 22:35:24.98 & +75:17:12.17 & 31.83 (0.003) & 40.53 & 3.3 $\times$ 2.7 (55.5) & 0.028\\
VLA 10 & 22:35:24.11 & +75:17:04.81 & 38.36 (0.004) & 53.12 & 3.3 $\times$ 2.8 (--30.6) & 0.037\\

\enddata

\end{deluxetable}

\clearpage

\begin{deluxetable}{ccccc}
\tabletypesize{\scriptsize}
\tablecaption{Bolometric Luminosity, Bolometric Temperature, and IR Spectral Indices of YSOs in L1251-C}
\tablewidth{0pt}
\tablehead{        \colhead{Sources} &
			\colhead{$L_{\rm{bol}}$} &
			\colhead{} &
			\colhead{$T_{\rm{bol}}$ (Class) } &
			\colhead{Spectral index $\alpha$ (Class)}\\
			\colhead{} &
			\colhead{[$L_{\sun}$]} &
			\colhead{} &
			\colhead{[K]} &
			\colhead{}}		
\startdata 
1 & 3.2 &  &622 (I)  & 2.94 (I)  \\
2 & 0.54 & &1180 (II) & --0.72 (II) \\
3 & 2.4 &  &1041 (II) & --0.12 (Flat) \\
4 & 3.1 &  &1074 (II) &  --0.51 (II) \\
5 & 0.25 &  &813 (II) & --0.38 (II) \\
6 & 0.085 &  &1440 (II) & --0.84 (II) \\
7 & 0.17 &  &747 (II) & --0.29 (Flat) \\
8 & 0.18 &  &910 (II) & --0.59 (II) \\
9 & 0.12 &  &960 (II) & --0.54 (II)\\
10 & 0.042  & &1209 (II) & --1.02 (II)\\
11 & 0.015  & &367 (I) & 0.35 (I) \\
12 & 0.47  & &1321 (II) & --0.84 (II) \\
13 & 0.63  & &1422 (II) & --1.05 (II)\\
14 & 0.16  & &713 (II) & --0.34 (II) \\
15 & 0.24  & &1441 (II) & --0.91 (II) \\
16 & 0.004  & &174 (I) & 1.20 (I) \\
17 & 0.80 & &763 (II) & 0.10 (Flat) \\
18 & 0.35 & &1505 (II) & --0.49 (II) \\
19 & 0.014  & &1537 (II) & --1.13 (II) \\
\enddata

\end{deluxetable}

\begin{figure}[!p]
\begin{center}
\plotone{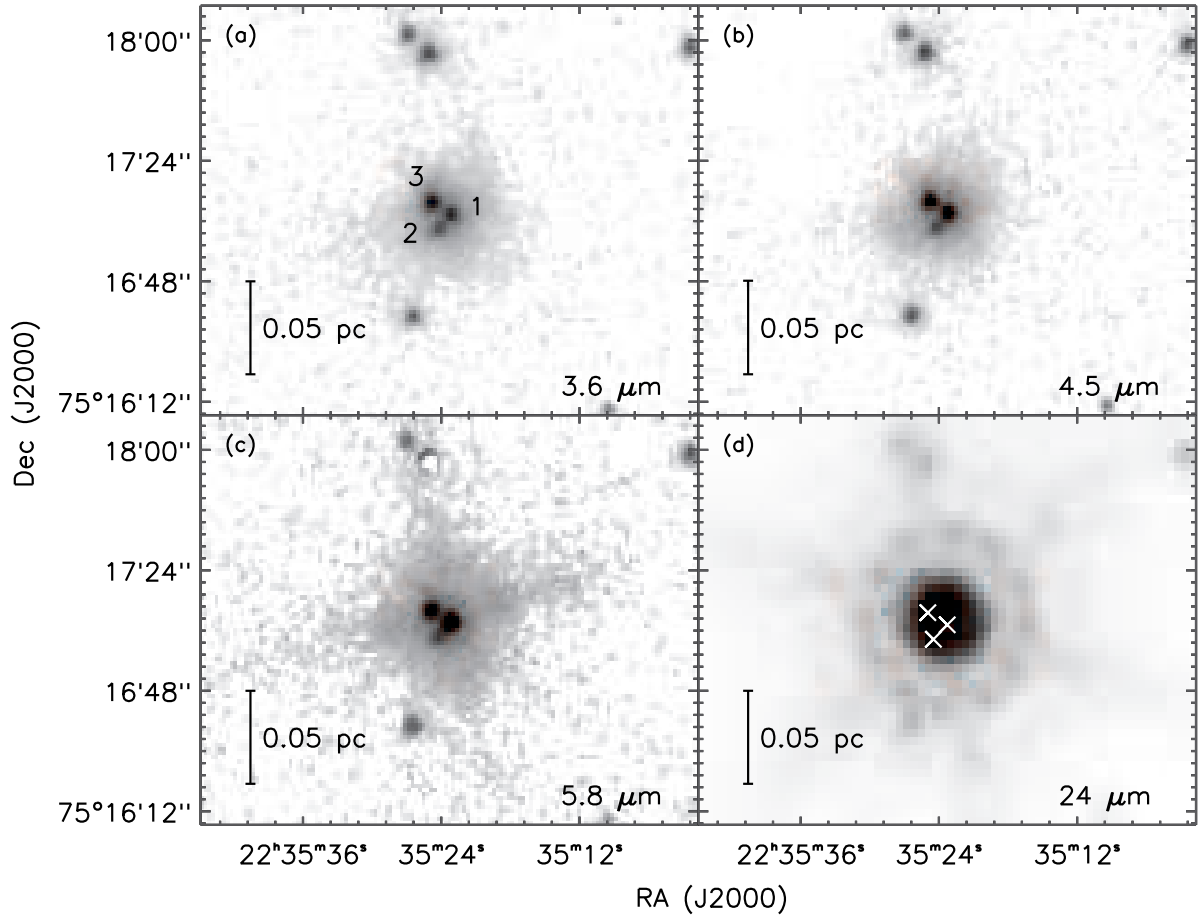}
\caption{The \textit{Spitzer} images of L1251A in log scale at (a) IRAC 3.6 \m, (b) IRAC 4.5 \m, (c) IRAC 5.8 \m, and (d) MIPS 24 \m. In Figure 1d, Xs indicate the positions of the IR sources in L1251A. }
\end{center}
\end{figure}

\clearpage

\begin{figure}[!p]
\begin{center}

\plotone{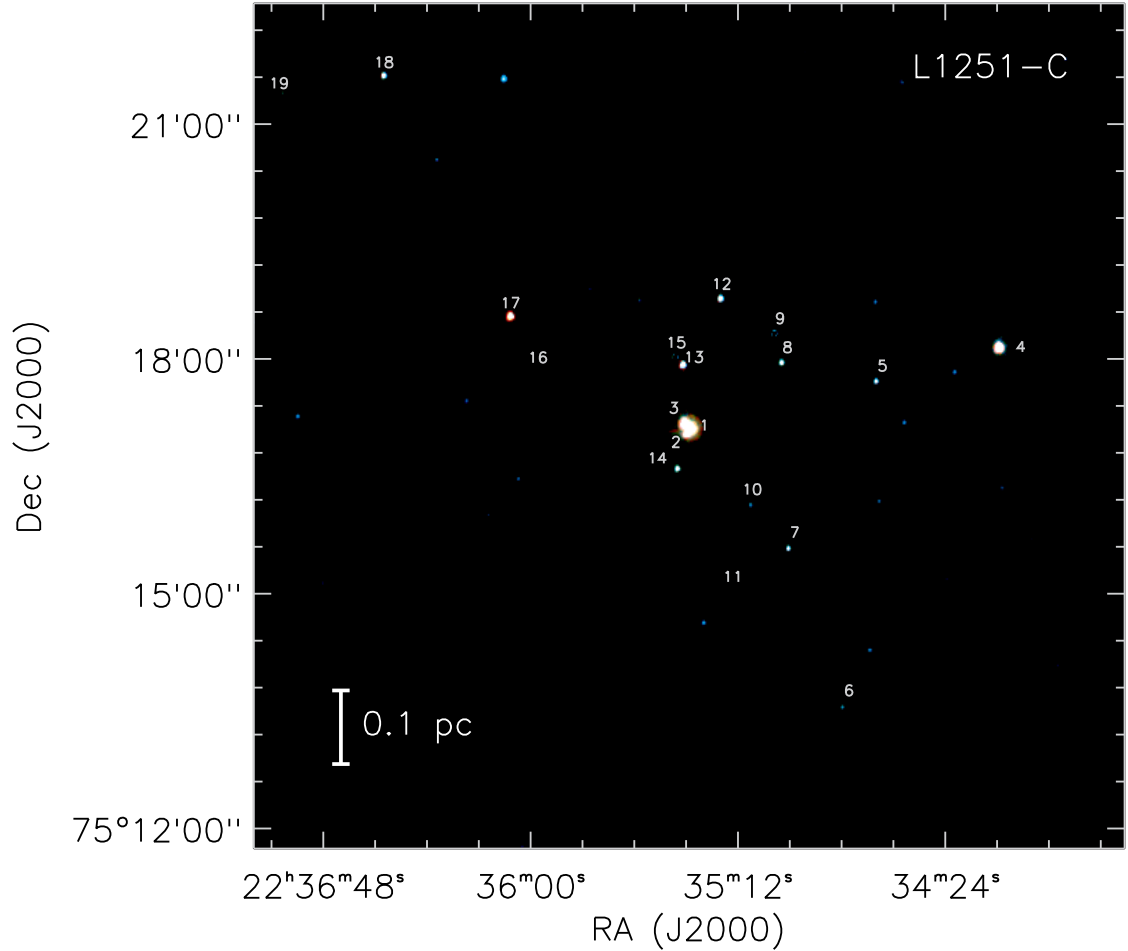}
\caption{The \textit{Spitzer} three-color composite image of L1251-C. The IRAC 3.6, 4.5, and 8.0 \m\ are presented in blue, green, and red, respectively. Numbers (4 to 19) are assigned along with Right Ascension after the numbers (1 to 3) of three sources in L1251A, which are also assigned along with Right Ascension.  }
\end{center}
\end{figure}

\clearpage

\begin{figure}[!p]
\begin{center}
\plotone{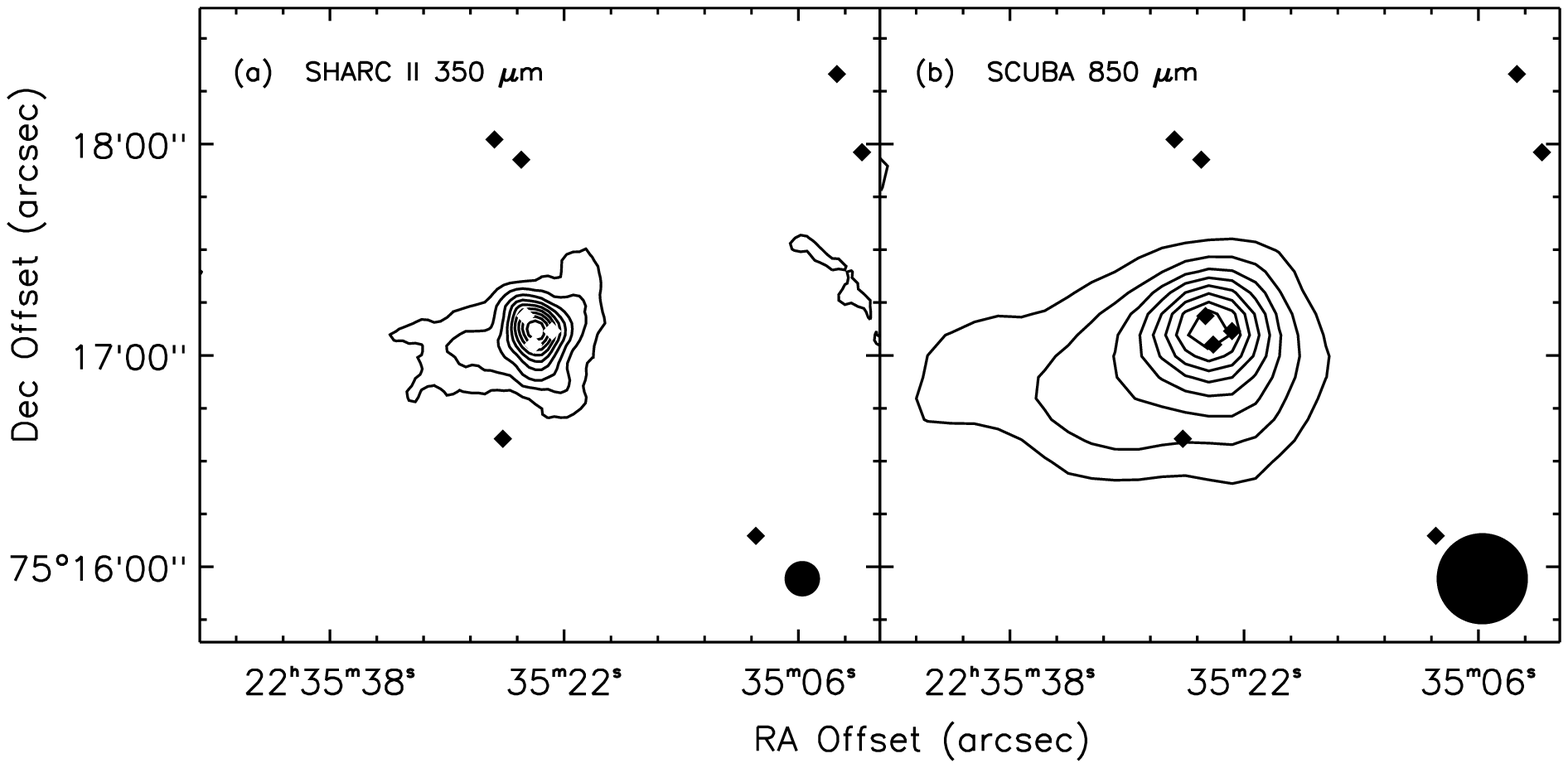}
\caption{Dust continuum emission map toward L1251A. (a) Continuum emission at 350 \m\ obtained with SHARC-II at the CSO \citep{wu07}. Contour levels  are 20, 40, 60, 80, and 100 $\%$ of the peak intensity, 5.40 Jy beam$^{-1}$. (b) Continuum emission at 850 \m\ obtained with SCUBA at the JCMT \citep{di08}. Contour levels are 20, 40, 60, 80, and 100 $\%$ of the peak intensity, 1.20 Jy beam$^{-1}$. The respective beam FWHM is presented in each panner with filled circle. 
YSO candidates are denoted by filled diamonds.}
\end{center}
\end{figure}

\clearpage

\begin{figure}[!p]
\begin{center}
\epsscale{0.95}
\plotone{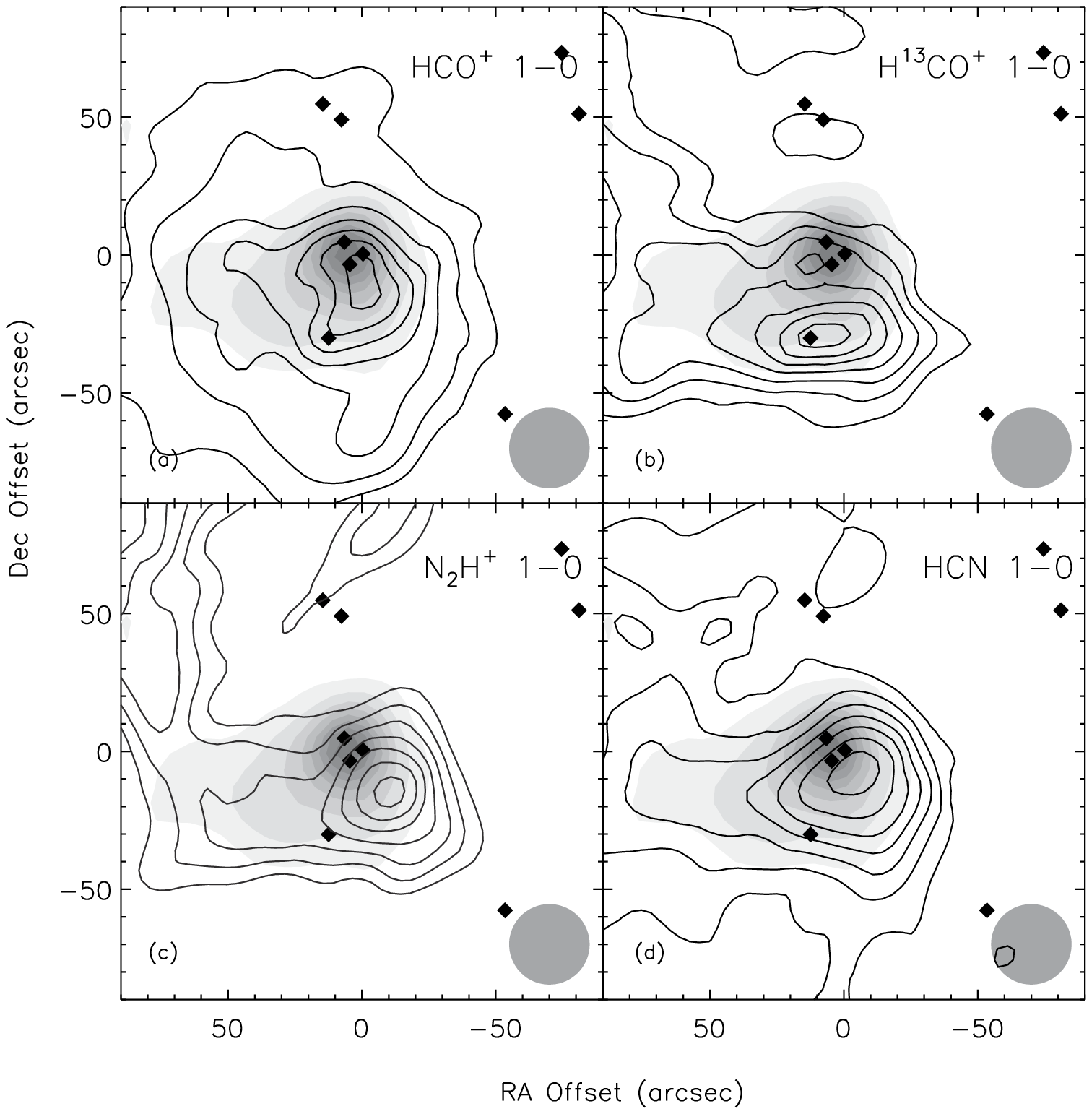}
\caption{Integrated intensity maps of (a) \hcop, (b) \htcop, (c) \nhp, and (d) \hcn\ on top of the 850 \m\ map (gray scale). 
Contour intervals are 10 $\%$ of the each peak intensity and range from 35 $\%$ to 95 $\%$.
The gray scale levels are 20, 40, 60, 80, and 100 $\%$ of the peak intensity, 1.20 Jy beam$^{-1}$.
YSO candidates are denoted by filled diamonds.
The filled circle at the bottom-right of each panel denotes the respective beam FWHM.} 

\end{center}
\end{figure}

\clearpage

\begin{figure}
\centering
\hspace{-0.8cm}
\includegraphics[width=0.4\paperwidth]{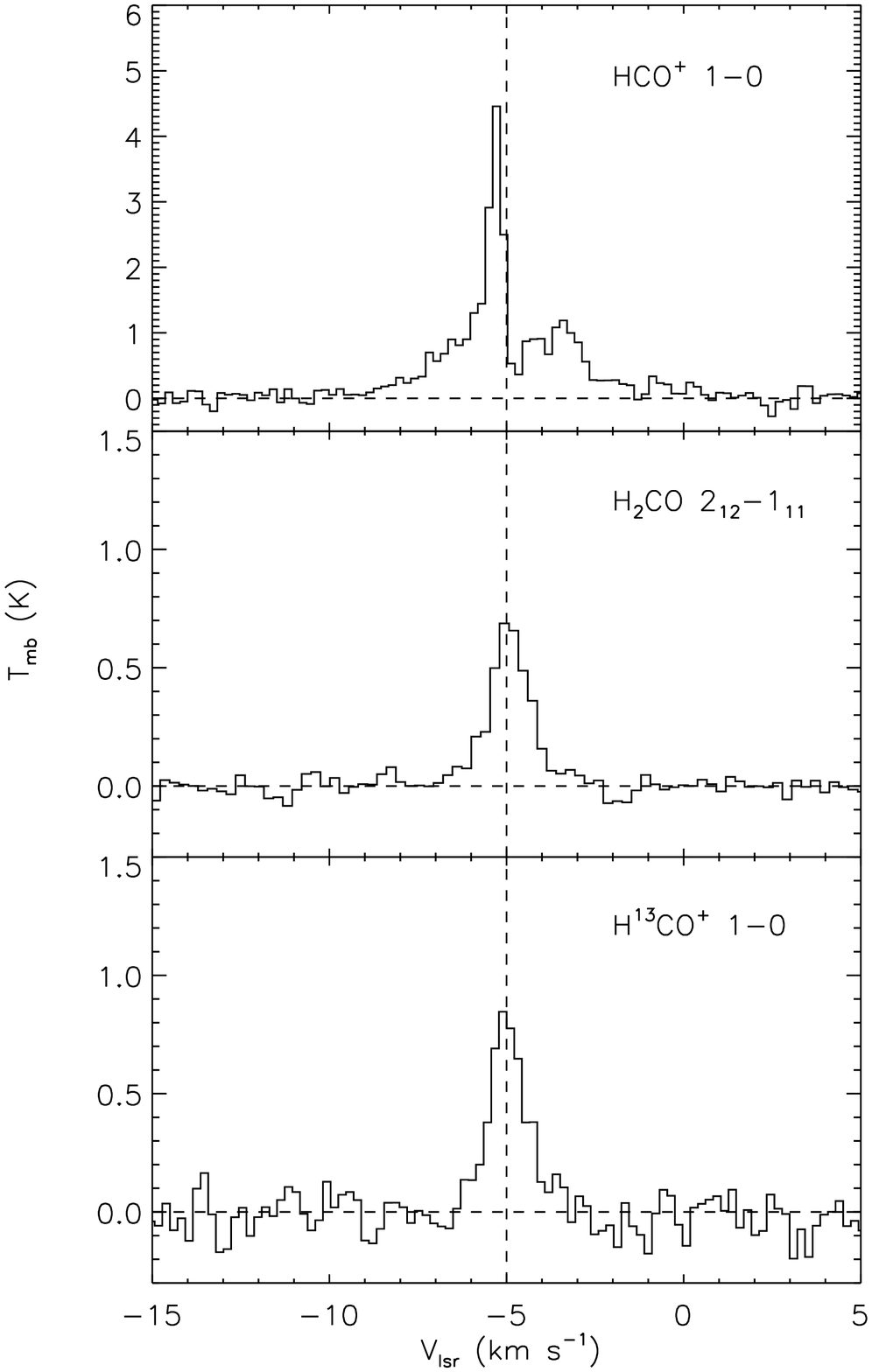}
\hspace{-0.5cm}
\includegraphics[width=0.4\paperwidth]{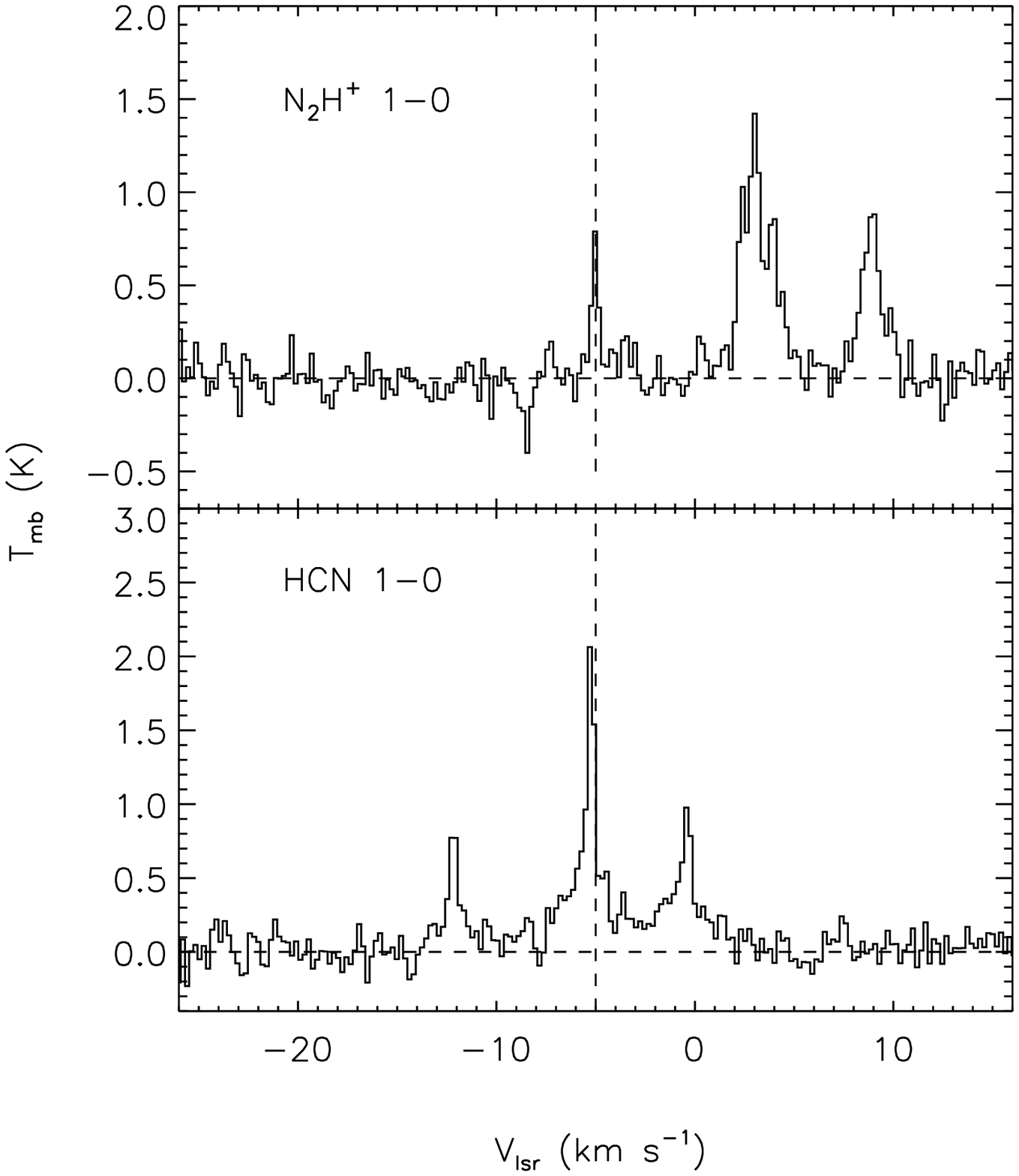}
\hspace{-1.0cm}
\caption{Spectra of the all molecular transitions observed with KVN at the center position of single-dish molecular line map. The horizontal dashed lines and the vertical dashed lines represent the zero baseline and the systemic velocity (--5 km s$^{-1}$) of L1251-C, respectively.}
\end{figure}

\clearpage

\begin{figure}[!p]
\begin{center}
\epsscale{0.95}
\plotone{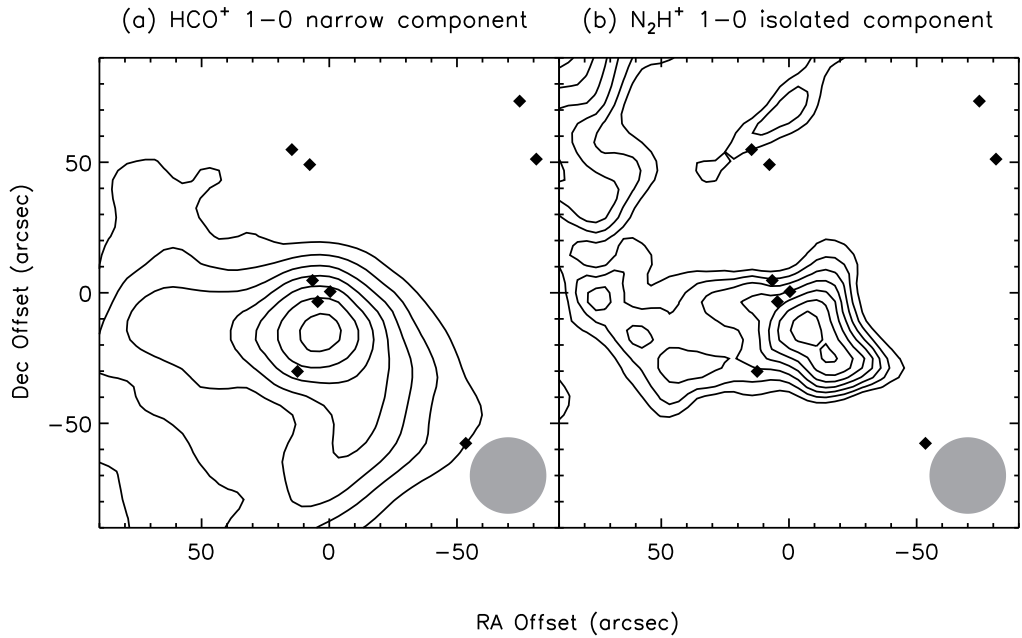}
\caption{Integrated intensity maps of (a) --5.3 km s$^{-1}$ (narrow) component of \hcop\ and (b) --5 km s$^{-1}$ (isolated) component of \nhp.  
YSO candidates are denoted by filled diamonds.
Contour intervals are 10 $\%$ of the each peak intensity and range from 35 $\%$ to 95 $\%$ .
The filled circle at the bottom-right of each panel denotes the respective beam FWHM. 
} 

\end{center}
\end{figure}

\clearpage

\clearpage

\begin{figure}[!p]
\begin{center}
\epsscale{0.5}
\plotone{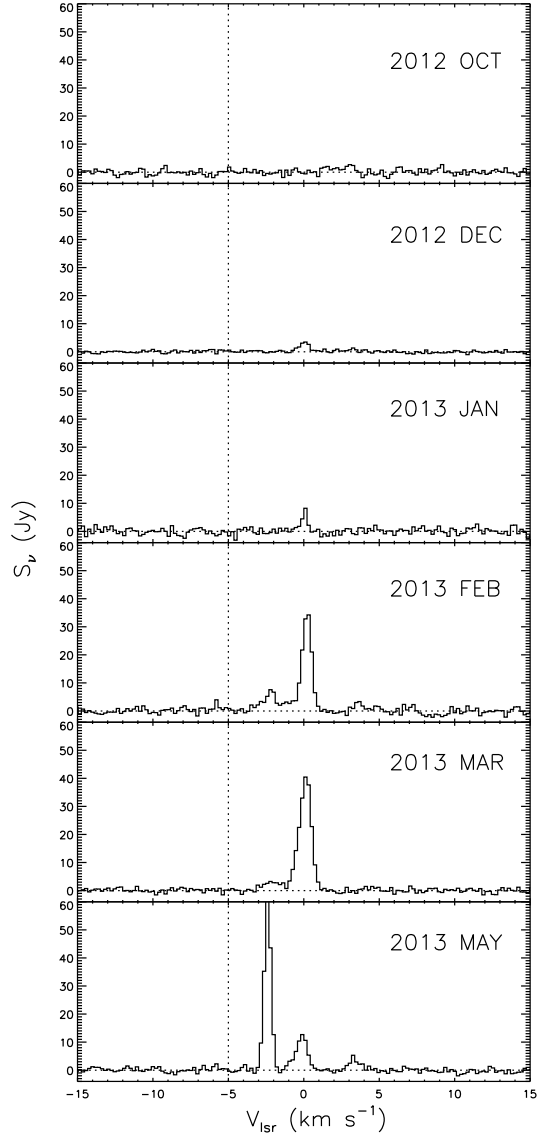}
\caption{Spectra of the 22 GHz \water\ maser line toward L1251A. The observation dates are labeled. The horizontal dotted line and the vertical dotted line represent the zero baseline and the systemic velocity (--5 km s$^{-1}$) of L1251-C, respectively. } 
\end{center}
\end{figure}

\clearpage

\begin{figure}[!p]
\begin{center}
\epsscale{1.0}
\plotone{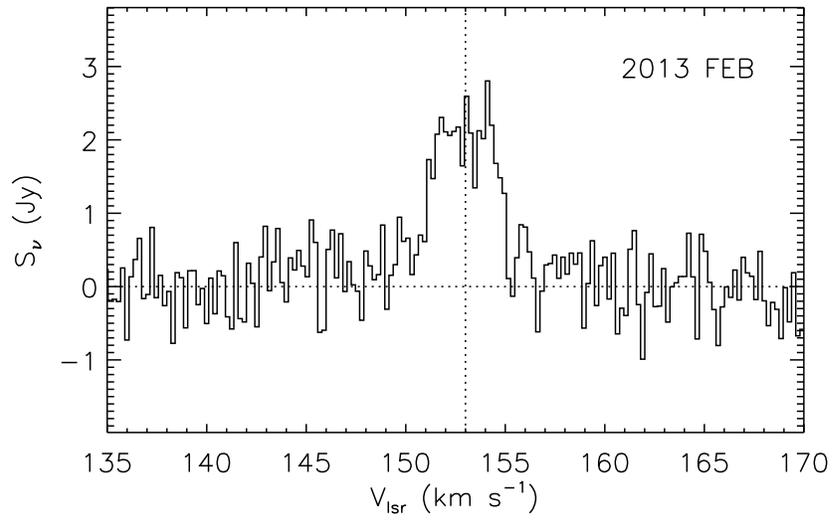}
\caption{The highest velocity component of 22 GHz \water\ maser line toward L1251A. This high velocity component was detected only on 2013 Feb. The horizontal dotted line and the vertical dotted line represent the zero baseline and the central velocity of the highest velocity component, respectively.}
\end{center}
\end{figure}

\clearpage

\begin{figure}[!p]
\begin{center}

\includegraphics[scale=0.7,angle=-90]{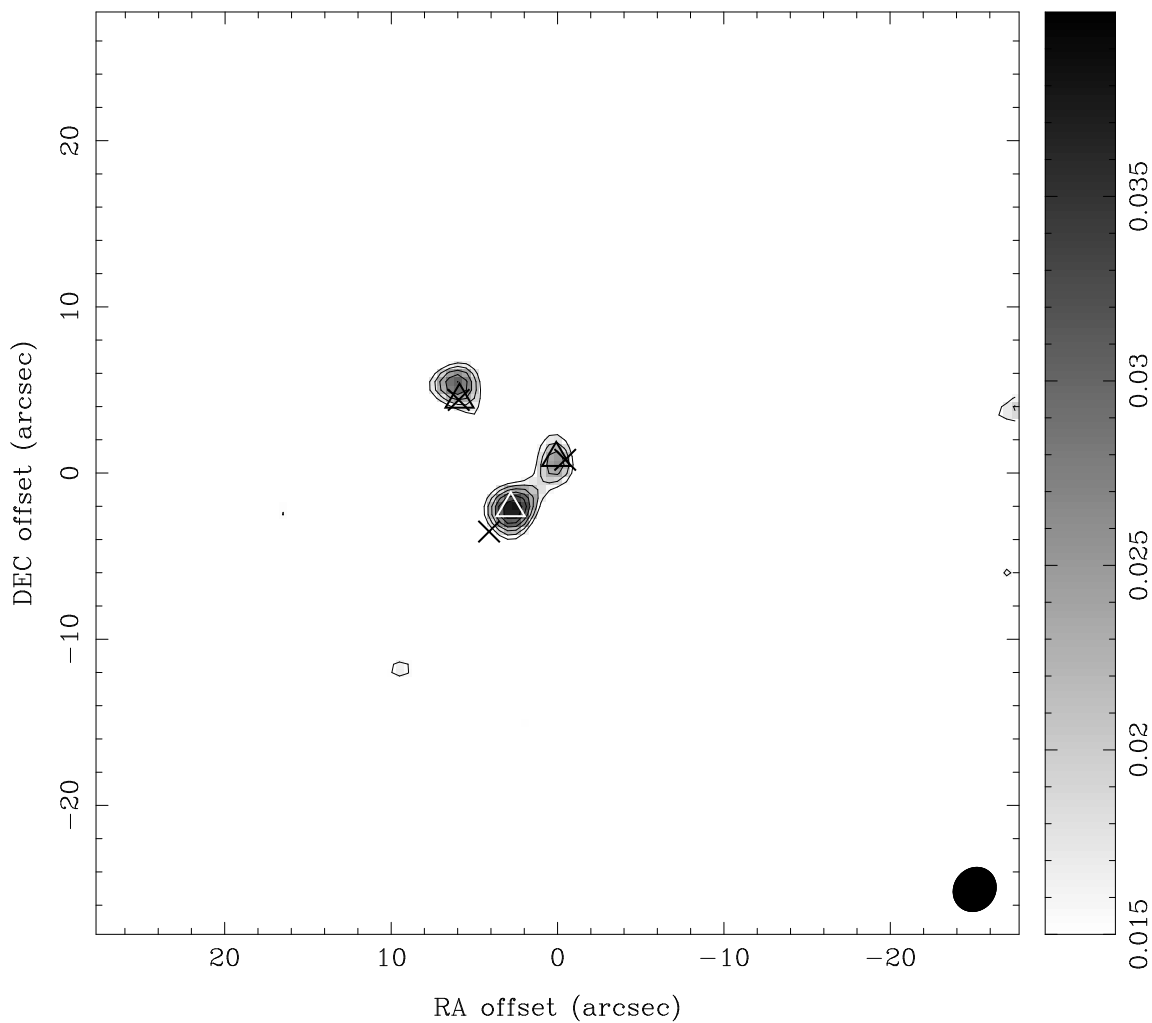}
\caption{1.3 mm Dust continuum emission map toward L1251A obtained with SMA. Contour levels are 40, 50, 60, 70, 80, and 90 $\%$ of the peak intensity, 0.038 Jy beam$^{-1}$ and the wedge indicates the continuum intensity scale in units of Jy beam$^{-1}$. The synthesized beam FWHM is presented at the bottom-right corner. The rms noise level 1 $\sigma$ is 4.7 mJy beam$^{-1}$.
The positions of IR sources in L1251A and the position of  VLA 3.6 cm continuum sources \citep{reipurth04} are denoted by Xs and triangles, respectively. The size of this presented map is the same as the primary beam size at this frequency ($\sim$55 $\arcsec$).  }
\end{center}
\end{figure}

\clearpage

\begin{figure}[!p]
\begin{center}
\epsscale{0.75}
\plotone{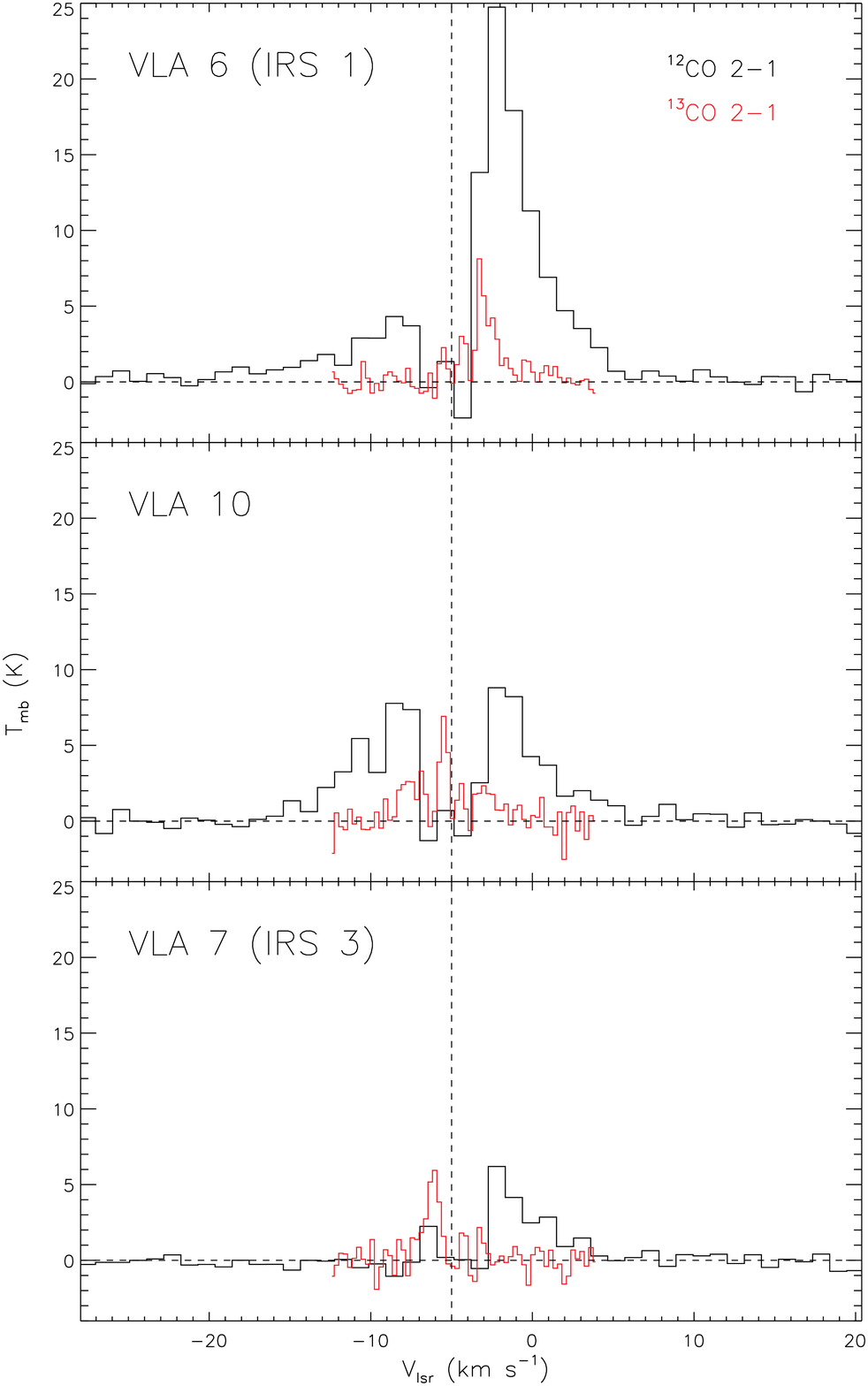}
\caption{Spectra of \tcot\ (black) and \thcot\ (red) observed with SMA toward the three SMA 1.3 mm continuum sources, VLA 6, VLA 10 and VLA 7, from the top to bottom. The black dotted vertical lines indicate the systemic velocity of L1251-C, --5.0 km s$^{-1}$, obtained from single dish observations.}
\end{center}
\end{figure}

\clearpage

\begin{figure}[!p]
\begin{center}
\includegraphics[angle=-90]{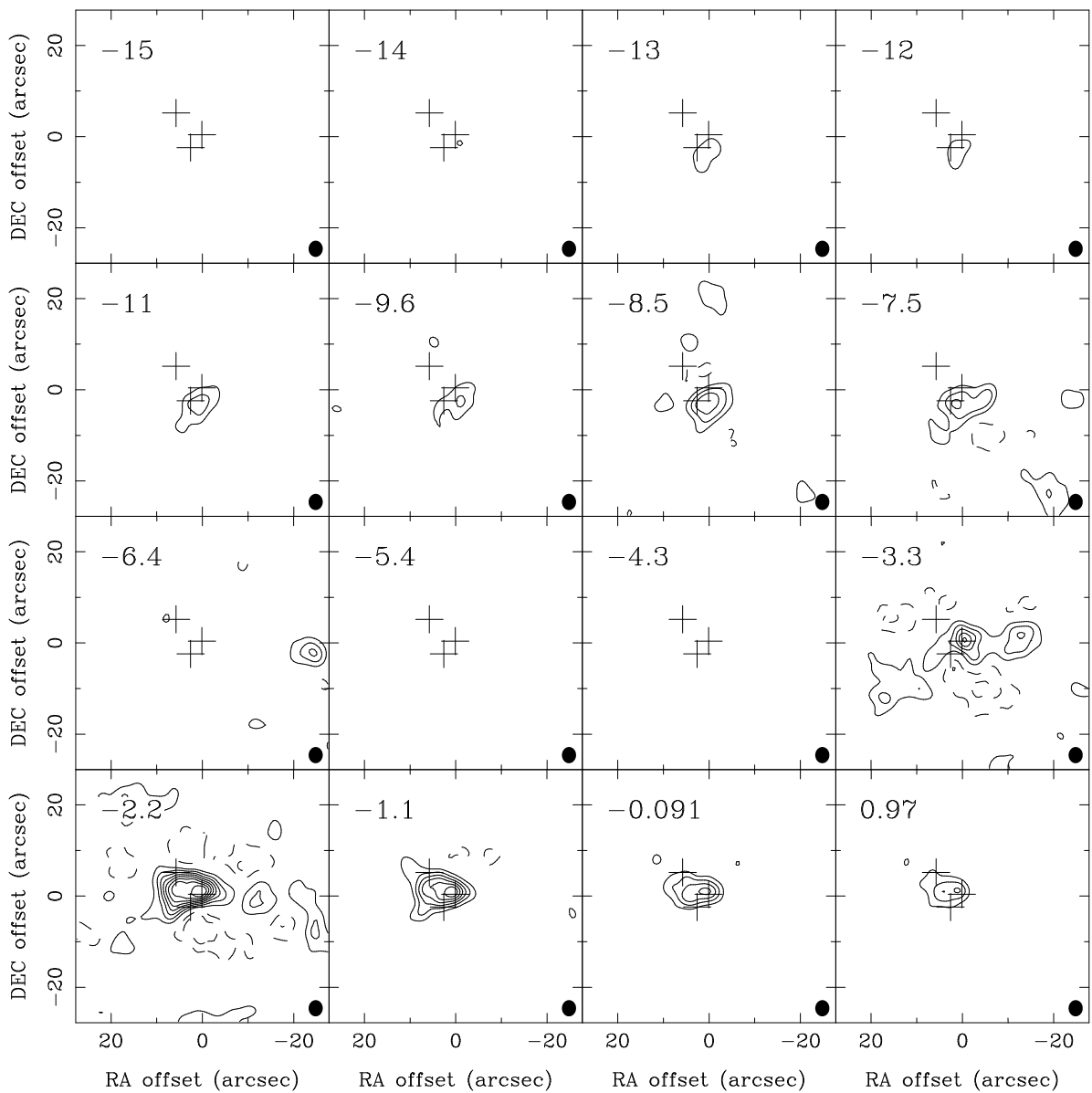}
\caption{Channel map of \tcot\ observed with the SMA. 
Respective velocities in \kms\ are indicated in the upper left of each panel. 
The contours start from 5 $\sigma$ and increase in steps of 5 $\sigma$, where 1 $\sigma$ is 0.268 Jy beam$^{-1}$. Dashed contours denote negative intensities with same steps of positive contours. Crosses denote the SMA 1.3 mm continuum source positions (VLA 6, VLA 7, and VLA 10). The synthesized beam FWHM is shown with the black ellipse.}
\end{center}
\end{figure}

\clearpage

\begin{figure}[!p]
\begin{center}
\epsscale{0.5}
\includegraphics[scale=0.7,angle=-90]{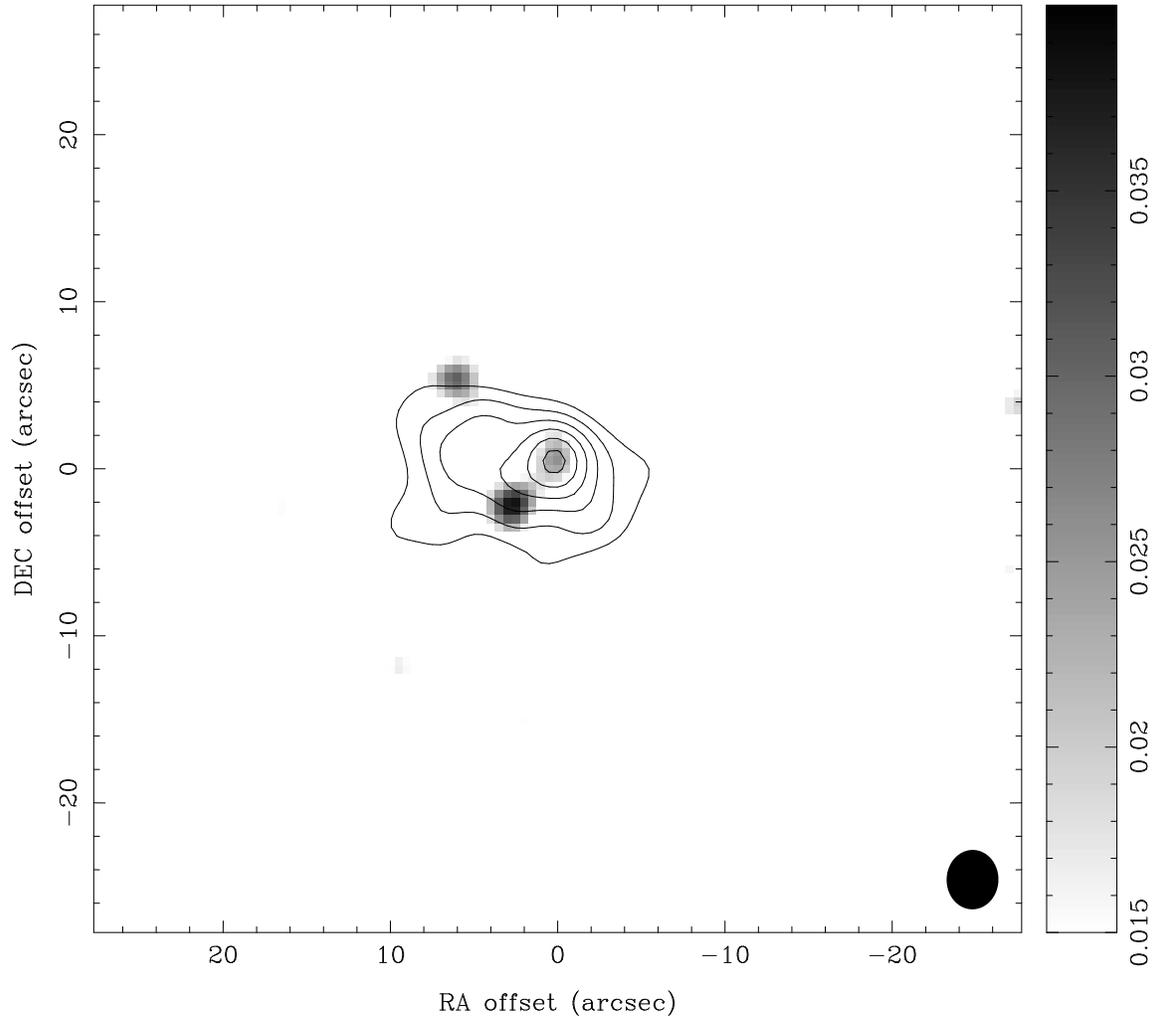}
\caption{Integrated intensity map of \tcot\ overlaid on the 1.3 mm continuum emission (gray scale). The contours represent 30, 45, 60, 75, 80, and 95 $\%$ of the peak intensity of 57.3 Jy beam$^{-1}$ \kms\ and the wedge indicates the continuum emission scale from 0.015 to 0.040 Jy beam$^{-1}$ . The synthesized beam FWHM is shown with the black ellipse. The map covers the primary beam size at this frequency ($\sim$55 $\arcsec$). }
\end{center}
\end{figure}

\clearpage

\begin{figure}[!p]
\begin{center}
\includegraphics[scale=0.7,angle=-90]{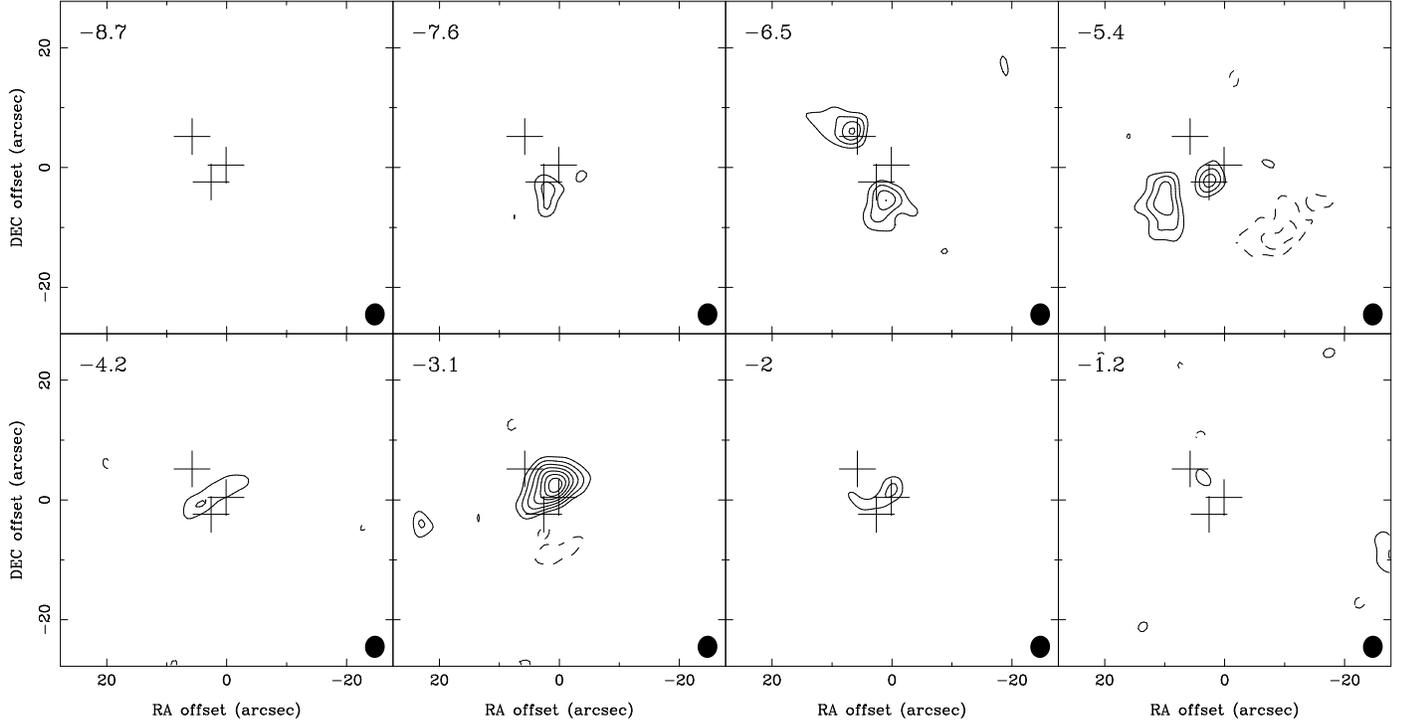}
\caption{Channel map of \thcot\ observed with the SMA. 
Respective velocities in \kms\ are indicated in the upper left of each panel. 
The contours start from 2 $\sigma$ and increase in steps of 1 $\sigma$, where 1 $\sigma$ = 0.404 Jy beam$^{-1}$. Dashed contours show the negative intensities with same steps of positive contours. The synthesized beam FWHM is shown with the black ellipse.}
\end{center}
\end{figure}

\clearpage

\begin{figure}[!p]
\begin{center}
\includegraphics[scale=0.7,angle=-90]{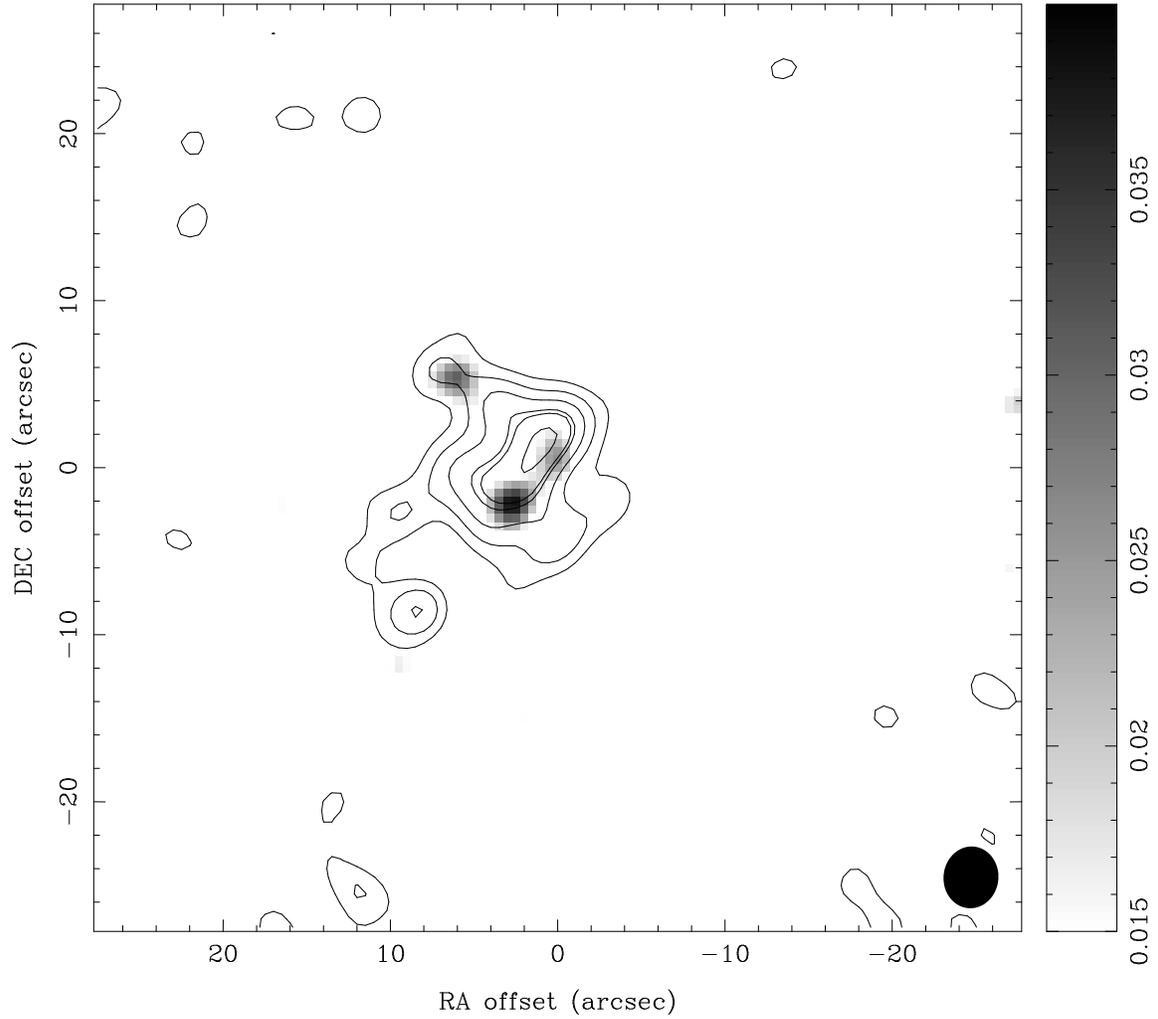}
\caption{Integrated intensity map of \thcot\ overlaid onto the 1.3 mm continuum (gray scale). The contours represent 25, 40, 55, 70, 85, and 100 $\%$ of the peak intensity of 7.75 Jy beam$^{-1}$ \kms\ and the wedge indicates the continuum emission scale from 0.015 to 0.040 Jy beam$^{-1}$. The synthesized beam FWHM is shown with the black ellipse. }
\end{center}
\end{figure}

\clearpage

\begin{figure}[!p]
\begin{center}
\epsscale{0.8}
\plotone{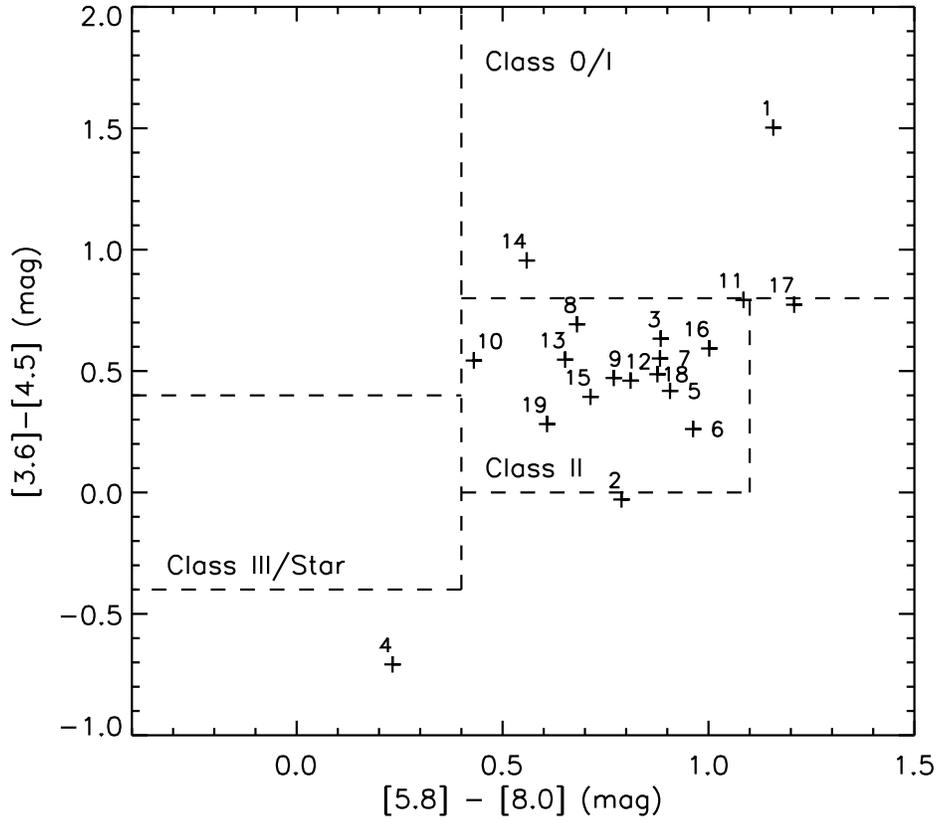}
\caption{[3.6] - [4.5] versus [5.6] - [8.0] color-color diagram for L1251-C YSO candidates in the IRAC bands. Source numbers are labeled. The dashed lines describe the approximate criteria for Class I, Class II, and Class III sources by \citet{lee06} based on \citet{allen04}.}
\end{center}
\end{figure}

\begin{figure}[!p]
\begin{center}
\plotone{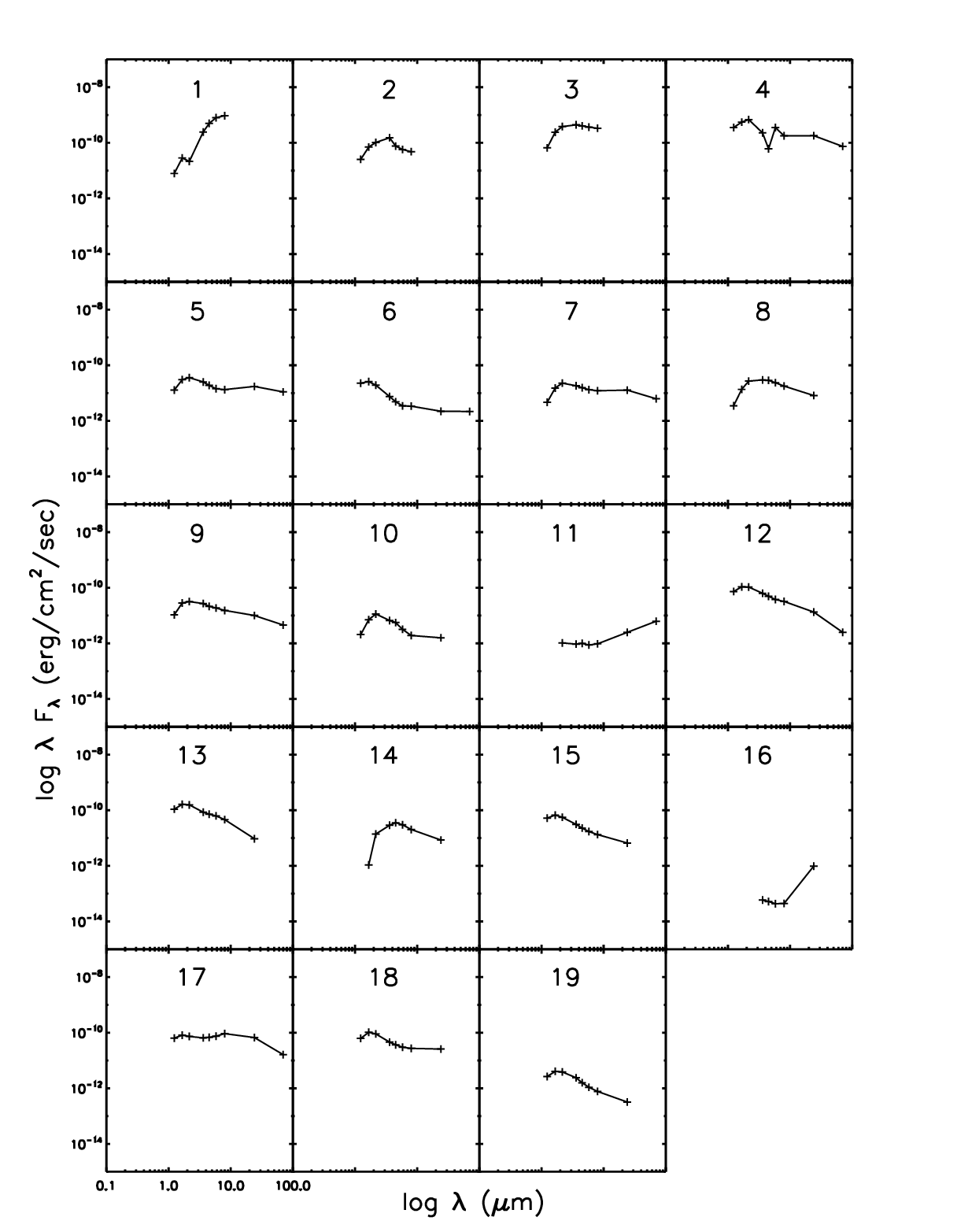}
\caption{SED of the YSOs in L1251-C. The black lines connect observed fluxes. }
\end{center}
\end{figure}

\clearpage

\begin{figure}
\centering
\hspace{-0.8cm}
\includegraphics[width=0.4\paperwidth]{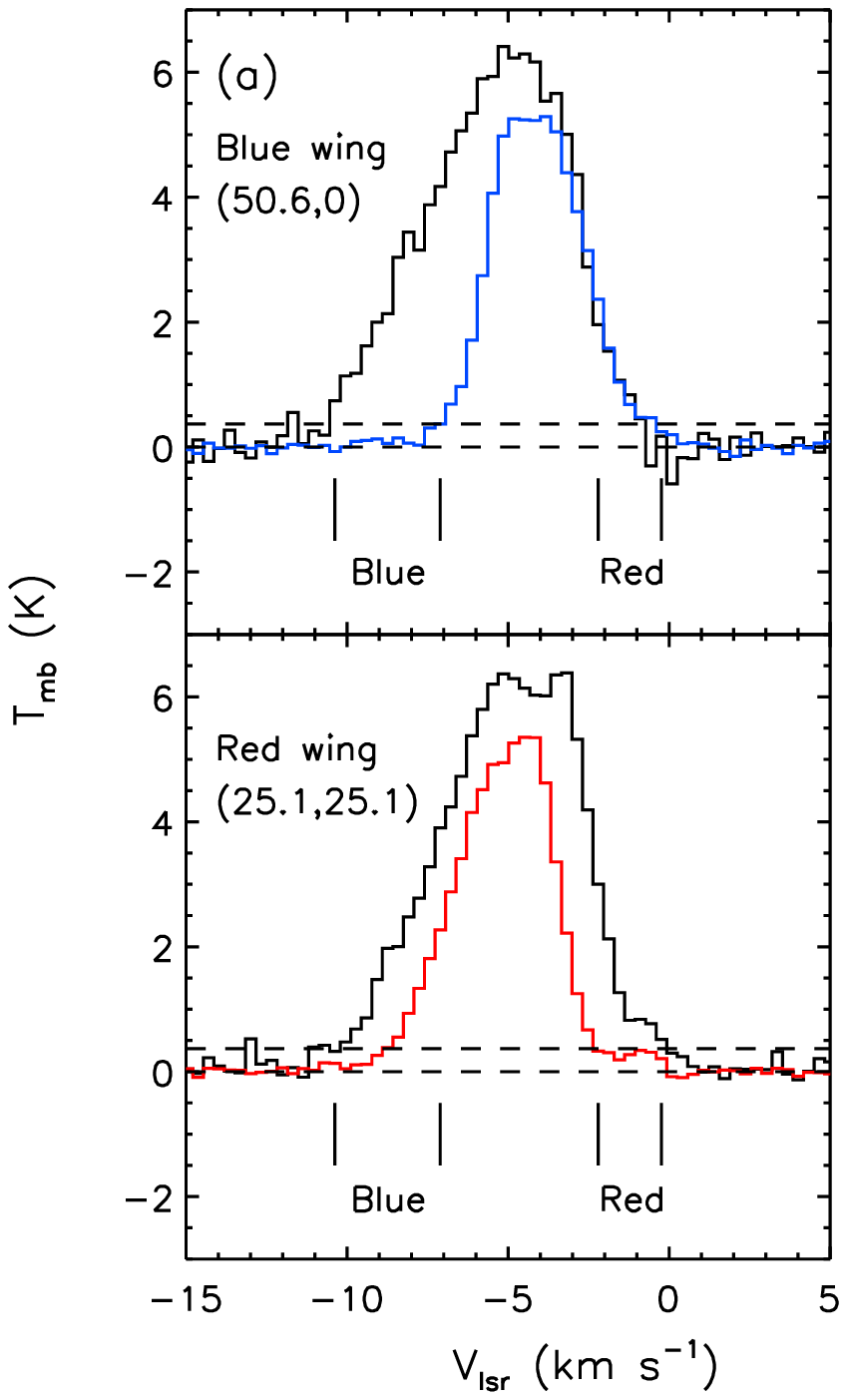}
\hspace{-2cm}
\includegraphics[width=0.45\paperwidth]{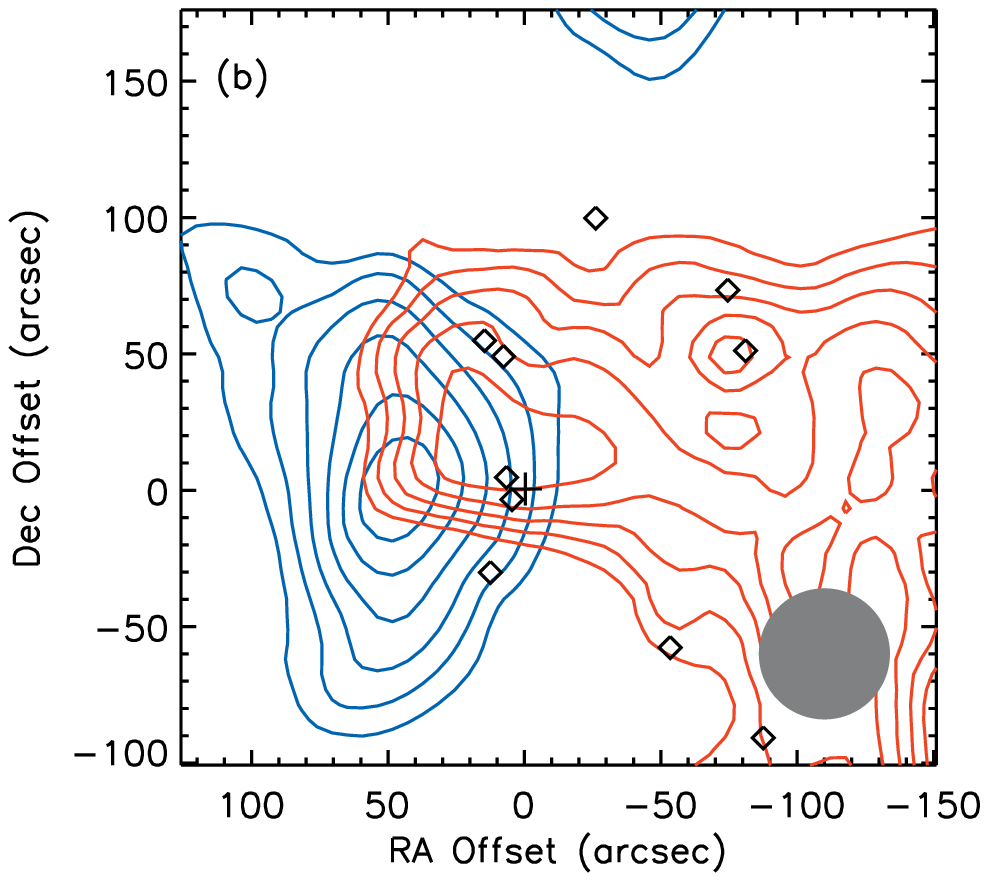}
\hspace{-1.0cm}
\caption{(a) The blue averaged spectrum of the blue wing free positions overlaid on the spectrum at the blueshifted emission peak position (\textit{top}) and the red averaged spectrum of the red wing free positions overlaid on the spectrum at the redshifted emission peak position (\textit{bottom}). (b) Integrated intensities of blue- and redshifted emission of \tco. Blue contours represent intensities integrated over the blue-shifted range from systemic velocity. Red contours show intensities integrated over the red-shifted range. The velocity range used for the blue and red components are from --10.4 to --7.1 km s$^{-1}$ and from --2.2 to --0.2 km s$^{-1}$, respectively. The contours start from 40 $\%$ of the peak intensity to 90 $\%$ and increase in steps of 10 $\%$. IRS 1 and the other YSOs are denoted by a cross and diamonds, respectively. The beam FWHM is presented with gray filled circle.}

\end{figure}

\clearpage

\begin{figure}[!p] 
\begin{center}
\plotone{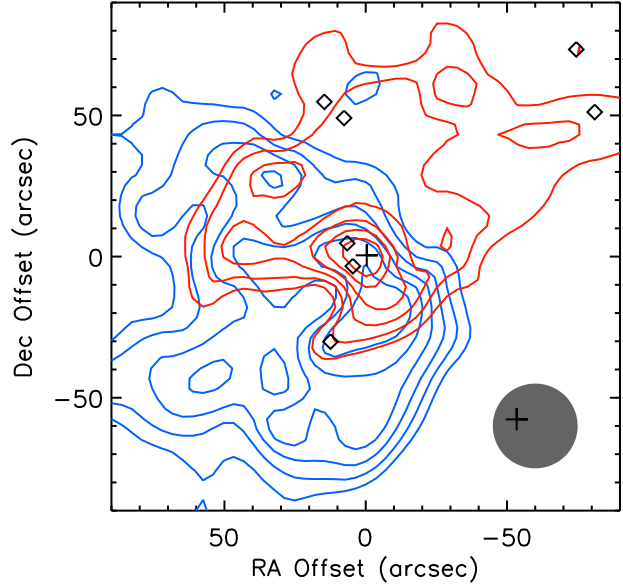}
\caption{Integrated intensities of blue- and redshifted emission of \hcop. Blue contours represent intensities integrated over the blueshifted range from systemic velocity. Red contours show intensities integrated over the redshifted range. The velocity range used for the blue and red components are from --9.7 to --5.9 km s$^{-1}$ and from --4.0 to --0.2 km s$^{-1}$, respectively. The contours start from 40 $\%$ of the peak intensity to 90 $\%$ and increase in steps of 10 $\%$. IRS 1 and the other YSOs are denoted by a cross and diamonds, respectively. The beam FWHM is presented with gray filled circle.}
\end{center}
\end{figure}

\clearpage

\begin{figure}[!p]
\begin{center}
\plotone{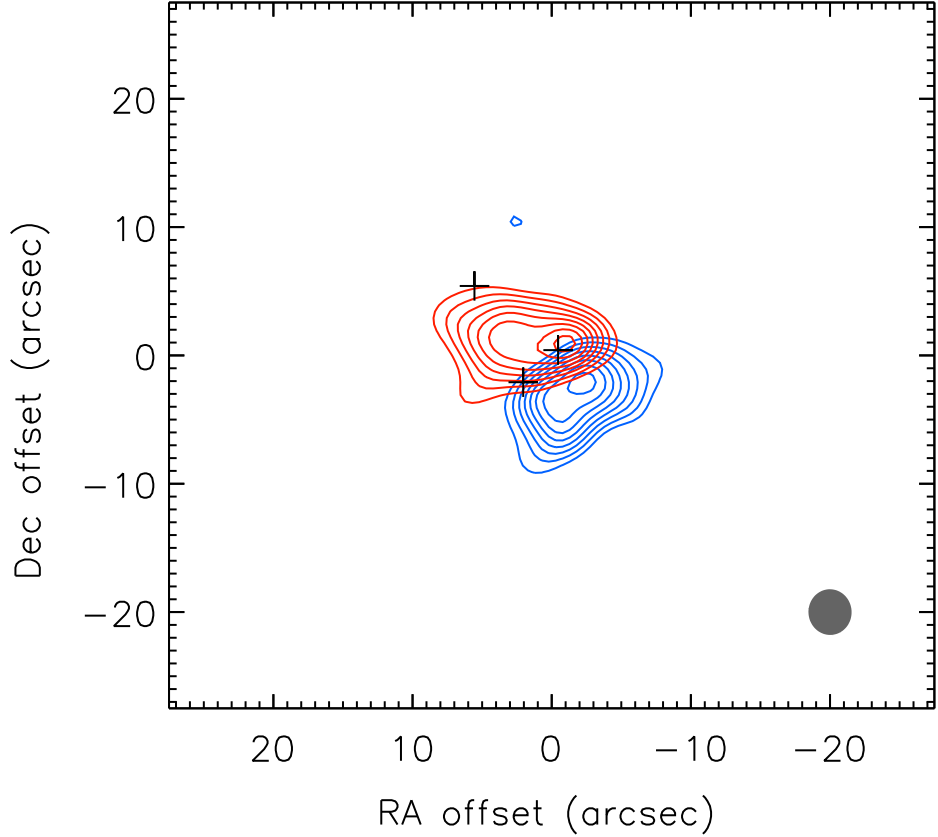}
\caption{Integrated intensities of blue- and redshifted emission of \tcot. Blue contours represent intensities integrated over the blueshifted range from --20 to --7 km s$^{-1}$ and red contours show intensities integrated over the redshifted range from --2 to 10 km s$^{-1}$. The contours start from 4$\sigma$ of the peak intensity to 90 $\%$ and increase in steps of 10 $\%$. VLA 6, VLA 7, and VLA 10 are denoted by crosses. The beam FWHM is presented with gray filled ellipse.}
\end{center}
\end{figure}

\end{document}